\documentclass[journal]{IEEEtran}
\usepackage{amsmath,amsfonts,amsthm,amssymb,bm}
\usepackage{algorithmic}
\usepackage{array}
\usepackage{setspace}
\usepackage[caption=false,font=normalsize,labelfont=sf,textfont=sf]{subfig}
\usepackage{textcomp}
\usepackage{stfloats}
\usepackage{url}
\usepackage{verbatim}
\usepackage{graphicx}
\usepackage{bbding}
\usepackage{multirow}
\usepackage{cite, bbm, xcolor}
\usepackage{refcount}
\usepackage{flafter}
\usepackage{makecell}
\usepackage{diagbox}
\usepackage{booktabs}
\usepackage{algorithm}

\newtheorem{lemma}{Lemma}

\newtheorem{remark}{Remark}
\theoremstyle{definition}

\usepackage{float}
\def\BibTeX{{\rm B\kern-.05em{\sc i\kern-.025em b}\kern-.08em
    T\kern-.1667em\lower.7ex\hbox{E}\kern-.125emX}}
\usepackage{balance}

\begin{document}

\title{Revealing computation-communication trade-off in Segmented Pinching Antenna System (PASS)}

\author{
        Deqiao Gan, Xiaoxia Xu, Xiaohu Ge, \textit{Senior Member, IEEE}, and Yuanwei Liu, \textit{Fellow, IEEE}
        \thanks{D. Gan and X. Ge (Corresponding author) are with the School of Electronic Information and Communications, Huazhong University of Science and Technology, Wuhan 430074, Hubei, China. (e-mail: gandeqiao@hust.edu.cn, xhge@mail.hust.edu.cn).}
        \thanks{X. Xu is with the School of Electronic Engineering and Computer Science, Queen Mary University of London, London E1 4NS, U.K. (email: x.xiaoxia@qmul.ac.uk).}
        \thanks{Y. Liu is with the Department of Electrical and Electronic Engineering, The University of Hong Kong, Hong Kong (e-mail: yuanwei@hku.hk).}
        \vspace{-1.3em}
}


\maketitle

\begin{abstract}
    A joint communication and computation (JCC) framework using segmented pinching antenna system (PASS) is proposed, where both the communication bit streams and computation data are simultaneously transmitted via uplink communications. The segmented PASS design is used to yield the tractable uplink transmission, and to mitigate large-scale path loss and in-waveguide loss.
    Based on three operating protocols, namely segment selection (SS), segment aggregation (SA), and segment multiplexing (SM), the joint transmit and receive beamforming problem is formulated: 1) The mean square error (MSE) minimization problem is formulated for computation-oriented cases. To address this problem, a low-complexity alternating optimization-minimum mean square error (AO-MMSE) algorithm is developed. This problem is decomposed into receiver-side and transmitter-side MSE subproblems that are iteratively optimized by MMSE receivers to obtain the closed-form solutions. It is mathematically proved that the segmented JCC-PASS framework significantly outperforms the conventional PASS for the average in-waveguide propagation gain. 2) The weighted sum rate (WSR) maximization problem is formulated for communication-oriented cases. To solve the decomposed receiver-side and transmitter-side MSE subproblems, the AO-weighted minimum mean square error (AO-WMMSE) algorithm is further developed. An auxiliary weight variable is introduced to linearize the WSR function and is alternatively optimized based on WMMSE to derive the closed-form solutions.
    Simulation results demonstrate that: i) The proposed JCC-PASS framework achieves up to 70.65$\%$ and 45.32$\%$ reductions in MSE compared with conventional MIMO and conventional PASS, and ii) it reaches 87.70$\%$ and 51.35$\%$ improvements in WSR compared with conventional MIMO and conventional PASS, respectively.
\end{abstract}

\begin{IEEEkeywords}
Computation-communication trade-off, pinching antenna system (PASS), weighted sum rate (WSR).
\end{IEEEkeywords}

\section{Introduction}
\IEEEPARstart{T}{he} computation-communication trade-off is a fundamental issue in 6G wireless networks, especially in emerging Internet of Things (IoT) and intelligent applications, where limited radio and computational resources must be efficiently coordinated to achieve low latency and high reliability \cite{yuanwei2023nearfield,wang2023isac,liu2022isac}. In conventional fifth-generation (5G) systems, the computation and communication are separately designed, where communication focuses on data transmission, computing relies on cloud/edge infrastructures, and sensing is often isolated from the communication links \cite{zhang2022isac6g}. Such functional isolation inevitably results in resource underutilization, long latency, and limited environmental awareness, which restricts the support of intelligent applications.
To achieve the integration of multi-functional communication and computation, the existing works have studied the mobile edge computing (MEC) and the integrated sensing, computation, and communication (ISCC).

The MEC enables partial or full offloading of computation tasks from user devices to nearby edge servers via uplinks, balancing communication and computation latency through adaptive resource scheduling \cite{mao2017mec, li2020mec, gan2024mec}. The energy-efficient joint communication and computation resource allocation for multi-user MEC systems was investigated by \cite{mao2017mec}, where minimizing total energy consumption leads to a trade-off between transmission rate and local processing time. Furthermore, the authors of \cite{li2020mec} introduced learning-assisted MEC mechanisms to adapt task allocation under dynamic wireless conditions, highlighting the need for intelligent cross-layer coordination. The FL-based framework for resource allocation in cloud-edge collaborative IoT systems was designed by \cite{gan2024mec}, which further captured the spatial-temporal coupling between communication and computation efficiency. It provides an effective means to reduce task delay and energy consumption in computation-intensive IoT applications.

To address these challenges, the concept of ISCC has been proposed as a key paradigm for 6G \cite{kaibin2025isac} to balance the communication and computation. The ISCC aims to make the trade-off in the computation and communication, which jointly optimizes transceiver design and computational task allocation, achieving unified utilization of spectrum, time, and spatial resources for synergistic multi-function operation \cite{qi2021iscc5g,qi2022iscc6g}. These works significantly reduce aggregation latency and improve spectral efficiency compared to conventional transmit-then-compute schemes. Building upon these advancements, the authors of \cite{qi2021iscc5g} proposed the AirComp architecture assisted by multiple-input multiple-output (MIMO), where the received signals from multiple devices are simultaneously aggregated over the air through analog waveform superposition, enabling fast global computation for IoT sensing data. This design reduces communication overhead but increases the distortion in aggregated results due to channel fading, reflecting an inherent tradeoff between communication reliability and computation accuracy. Then, the authors of \cite{qi2022iscc6g} developed multi-functional beamforming to jointly serve sensing and distributed computation tasks by exploiting the spatial degrees of freedom of large-scale antenna arrays. Although spatial multiplexing improves spectral efficiency, it inevitably introduces interference among concurrent functions, again highlighting the fundamental balance between communication capacity and computational performance in massive IoT scenarios.

However, existing MEC and ISCC realizations based on the massive MIMO systems are constrained by their rigid antenna structures and limited spatial flexibility. The fixed placement of antenna elements and the discrete control of reflecting elements hinder continuous beam adaptation and environment-aware reconfiguration, especially at high-frequency bands where line-of-sight (LoS) links and path-loss control are critical. Consequently, it is essential to develop a reconfigurable physical-layer platform that can dynamically reshape spatial channels to better support the joint operation of sensing, communication, and computation.

Conceptually, the pinching antenna system (PASS) serves as a new form of flexible antenna technology that offers greater adaptability and scalability. PASS was first prototyped by NTT DOCOMO \cite{2022NTTDOCOMO} and later introduced into wireless communications \cite{ding2024pass}. An earlier concept known as surface-wave communication superhighways was proposed in \cite{kit2021radio}, which leverages similar principles of in-waveguide propagation on reconfigurable surfaces. Different from the conventional reconfigurable antenna architectures, PASS employs dielectric and extensible waveguides embedded with separately configurable pinching antennas (PAs), which can be dynamically activated and repositioned along the waveguide \cite{yuanwei2025pass, gan2025NOMAPASS}. With reconfigurable waveguides and PAs for in-waveguide propagation, PASS can significantly reduce path loss and improve transmission efficiency \cite{liu2024path, chu2024propagation}. Since PASS enables the flexible deployment of PAs close to end users, it establishes robust LoS connections and effectively mitigates large-scale path loss while overcoming coverage gaps \cite{wang2025pass,xu2025pass2}. Hence, PASS realizes a novel form of pinching beamforming, where both the large-scale attenuation and the phase of the transmitted signal can be continuously reconfigured in the spatial domain \cite{ouyang2025array, xu2025antenna}.
In uplink transmission, PASS enables flexible activation of user-proximal and reconfigurable PAs, which significantly enhances uplink channel gain and spatial diversity, thereby reducing transmit power requirements and mitigating link blockage for energy-efficient and reliable multiuser uplink communication \cite{ouyang2025uplink}. This paper proposed the novel segmented waveguide-based PASS under three protocols to enable the multi-PA uplink transmission.

Benefiting from these properties, considering the computation-communication trade-off in PASS framework enables an intrinsically reconfigurable and spatially efficient architecture, capable of coordinating multi-functional wireless operations with high adaptability. The computation-communication trade-off is considered in ISAC and ISCC systems, and can be utilized to reveal the computation-communication performances for PASS, which has recently attracted growing attention for its potential to enable physical-layer reconfigurability.
However, current PASS studies mainly emphasize communication-oriented objectives, such as capacity enhancement, power transfer efficiency, and beamforming optimization \cite{ding2024pass, wang2025pass, ouyang2025array, gan2025NOMAPASS}, leaving the sensing-computation coordination unexplored.
Few works have addressed how PASS can facilitate multi-functional coordination among communication and computation, or how its unique spatial degrees of freedom can be systematically exploited for joint optimization. Most existing architectures adopt a monolithic waveguide configuration in which all PAs are simultaneously activated, leading to strong mutual coupling, high control overhead, and inefficient energy utilization.

Although the computation-communication trade-off of MIMO systems has been extensively studied, current computation-communication trade-off in PASS framework is still in its infancy and faces several fundamental challenges:
\begin{itemize}
    \item \emph{Mutual interference:} The uplink sensing and computation coordination remains largely unexplored. Because the lack of joint uplink transceiver optimization leads to severe mutual interference between sensing and computation signals, degrading the overall performance.
    \item \emph{Fully activated waveguide:} The most reported architectures adopt monolithic or fully activated waveguide structures, where all PAs are simultaneously activated. The configuration introduces strong intra-waveguide coupling and high control overhead, making it difficult to achieve fine-grained spatial reuse and low-complexity control. These designs restrict the scalability of PASS in large-area or multi-user deployments.
\end{itemize}

These open problems highlight the need for a segmented and computation-communication trade-off PASS architecture, capable of supporting flexible segment activation and distributed signal processing to realize fully integrated 6G functionalities. In this paper, we propose an uplink JCC-PASS framework, where the segmented waveguide-enabled PASS is deployed to simultaneously transmit the communication bits and computation data under three protocols.
This paper presents two optimization problems from computation and communication perspectives. Specifically, we formulate the computation-oriented mean square error (MSE) minimization problem to jointly optimize the transmit beamforming and receive beam at the user equipment (UE) and the base station (BS). Since this problem is highly nonconvex and coupled, we decompose it into receiver-side and transmitter-side MSE subproblems. Then, we develop the alternating optimization and the minimum mean square error (AO-MMSE) algorithm to obtain the closed-form solutions, which utilizes MMSE receivers to alternatively optimize the transmit beamforming and receive beams. For the communication-oriented weighted sum rate (WSR) maximization problem, we design the AO-weighted minimum mean square error (AO-WMMSE) algorithm to derive the closed-form solutions. The decomposed receiver-side and transmitter-side WSR subproblems are iteratively updated by the WMMSE. The key contributions are summarized as follows.

\begin{enumerate}
    \item We propose an uplink JCC-PASS framework, which transmits both computation data and communication bit streams to BS and mitigates the large-scale loss by the segmented PASS architecture. We formulate the MSE minimization problem for communication-oriented scenarios and WSR maximization problem for computation-oriented scenarios where the transmit and receive beamforming at UEs and the BS are jointly optimized. To handle the high convexity and couplings of these two problems, we respectively decompose them into nested receiver-side and transmitter-side subproblems.
    \item For the computation-oriented MSE minimization problem, we develop the AO-MMSE algorithm to jointly optimize the transmit and receive beamforming. Specifically, we first minimize the receiver-side MSE subproblem employing MMSE receivers to obtain the closed-form of receive beams under the fixed transmit beamforming. Based on the semi-definite relaxation (SDR) method, we convert the transmitter-side MSE subproblem into convex formulation with the above receive beams for the closed-form transmit beamforming. It is mathematically proved that the JCC-PASS framework has the superiority over the conventional PASS for the average in-waveguide propagation gain.
    \item For the communication-oriented WSR maximization problem, we design the AO-WMMSE algorithm to derive the closed-form solutions. We maximize the receiver-side WSR subproblem with the fixed transmit beamforming set by MMSE receivers for the closed-form receive beams. Further, we introduce the auxiliary weight variable to linearize the transmitter-side WSR subproblem and then derive the closed-form transmit beamforming, leading to a computationally efficient procedure with guaranteed monotonic improvement of the objective function.
    \item Extensive numerical results verify the effectiveness of the proposed JCC-PASS framework and developed algorithms, which demonstrates that: i) The proposed JCC-PASS framework decreases the MSE by 70.65$\%$ and 45.32$\%$ compared with conventional MIMO and conventional PASS, and ii) it achieves 87.70$\%$ and 51.35$\%$ improvements in WSR compared with conventional MIMO and conventional PASS, respectively.
\end{enumerate}

\textit{Notations}: $\mathbb{C}^{M \times 1}$ and $\mathbb{R}^{M\times 1}$ denote the space of $M \times 1$ complex valued vectors and real valued vectors, respectively. 
The variable, vector, and matrix are denoted by $x$, $\mathbf{x}$, and $\mathbf{X}$, respectively.
$|x|$ denotes the absolute value of a real number and the modulus of a complex number. The notation $\text{rank}(\mathbf{X})$ defines the rank of $\mathbf{X}$.
$\text{Re}\left\{x\right\}$ and $\text{Im}\left\{x\right\}$ denote the real and image parts of $x$, and $x^{H}$ is the complex conjugate number of $x$. 
$\mathbf{I}_{N\times 1}$ denotes an $N$-dimension all-ones vector.
$\mathbf{X}^{T}$ and $\mathbf{X}^{H}$ denote the transpose and the Hermitian matrix.

The remainder of the paper is organized as follows. Section II introduces the system model of the proposed JCC-PASS framework. Section III presents the AO-MMSE-based algorithm for the computation-oriented MSE minimization problems under three operating protocols. Section IV provides the AO-WMMSE-based algorithm for the communication-oriented WSR maximization problem. Section V demonstrates simulation results and analysis. Finally, Section VI concludes this paper.


\section{System Model}
\begin{figure}[!t]
    \centering
    \includegraphics[width=3.3in]{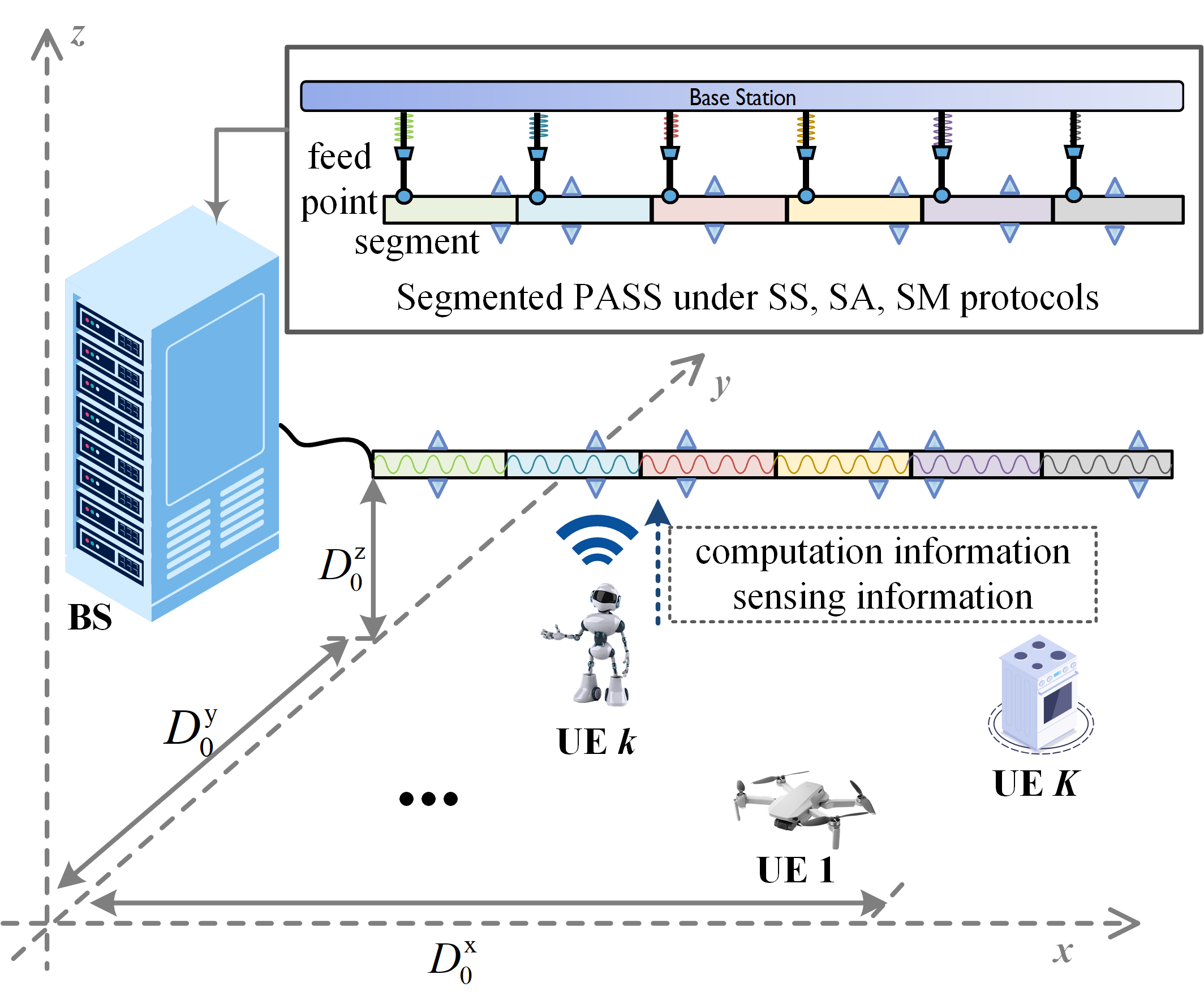}
    \caption{System model: The proposed JCC-PASS framework.}
    \label{systemmodel}
\end{figure}

Consider an uplink transmission assisted by PASS, where a BS equipped with a segmented waveguide. We propose a JCC-PASS framework to serve $K$ UEs with sensing and communication capabilities, shown as Fig. \ref{systemmodel}. As depicted in Fig. \ref{systemmodel}, each waveguide is divided into $M$ isolated segments, each equipped with one reconfigurable PA. This segmentation enables independent control of radiation points, forming the basis of the segmented PASS protocols. Different from the conventional PASS framework with a continuous waveguide, we consider the segmented waveguide-enabled PASS \cite{ouyang2025uplink}, where the dielectric waveguide is divided into $M$ short segments of equal length $L$ along the $x$-axis to connect with BS. These segments are electrically isolated and equipped with their own independent feed points, through which signals are fed into the waveguide and relayed to the BS via low-loss wired links, e.g., optical fiber. Because the wired links incur negligible attenuation compared with the dielectric propagation loss, they are assumed to be lossless.

Without loss of generality, the waveguides are deployed parallel to the $x$-axis at the height of $D^{\text{z}}_0$, with PAs pinched along the waveguide. Notably, as $M=1$, this segmented PASS is same as the conventional continuous waveguide-enabled PASS. We denote the feed point location of segment $m$ on the waveguide by $\bm{\psi}^{m}_0 = [{\psi}^{m}_0,D^{\text{y}}_0,D^{\text{z}}_0]$, where $D^{\text{y}}_0 \in [0,D^{\text{y}}_{\max}/2]$, $\psi^1_0 < \psi^2_0 < \dots < \psi^M_0 < ML$, and $m \in \mathcal{M}=\{1,2,\dots,M\}$. Let the locations of activated PAs $N_m \in \mathcal{N}=\{1,2,\dots,N_{\max}\}$ in the segment $m$ of the waveguide be ${\bm{\psi}}^{m}_{n}=[\psi^m_n, D^{\text{y}}_0, D^{\text{z}}_0]$ with the location constraints $\psi^m_0 \le \psi^m_n \le \psi^m_0 + L$ and $|{\psi}^{m}_{n'} - {\psi}^{m}_n| \ge \Delta, \forall n \ne n'$, where $\Delta > 0$ means the minimum inter-antenna spacing required to mitigate mutual coupling effects. Suppose that UE $k$ is in the rectangular area $D^{\text{x}}_0 \times 2D^{\text{y}}_0$ with the location $\bm{\psi}_k = [{\psi}^{\text{x}}_k, {\psi}^{\text{y}}_k, 0]$, where $k \in \mathcal{K} =\{1,2,\dots,K\}$, ${\psi}^{\text{x}}_k \in [0,ML]$, $D^{\text{x}}_0 = ML$ and ${\psi}^{\text{y}}_k \in [0, 2D^{\text{y}}_0]$.

\subsection{PASS Channel Model}
The segmented waveguide-enabled PASS is operated at high-frequency band where LoS propagation dominates, we adopt a free-space LoS model for the PA-UE link. The baseband channel coefficient between the PA $n$ of segment $m$ and UE $k$ can be expressed as
\begin{equation}
    \label{PA_UE_channel}
    h_o(\bm{\psi}_k, \bm{\psi}^m_n) = \frac{\eta {e}^{-j\kappa \|\bm{\psi}_k - \bm{\psi}^m_n\|}}{\|\bm{\psi}_k - \bm{\psi}^m_n\|}, k \in \! \mathcal{K} \! = \! \{1,2,\dots,K\},
\end{equation}
where $\kappa=\frac{2\pi}{\lambda}$ represents the wave-domain number, and $\lambda$ is the wavelength, and the constant $\eta=\frac{c}{4\pi f_c}$ depends on the speed of light $c$ and carrier frequency $f_c$. Within each segment, the in-waveguide propagation from the feed point $\bm{\psi}^m_0$ to the PA $N_m$ can be expressed as
\begin{equation}
    \label{PA_feed_channel}
    \begin{aligned}
    h_i(\bm{\psi}^m_n, \bm{\psi}^m_0) & = 10^{-\frac{\kappa_0}{20}\|\bm{\psi}_n - \bm{\psi}^m_0\|}e^{\frac{-j2\pi\|\bm{\psi}_n - \bm{\psi}^m_0\|}{\lambda_g}} \\
    & = 10^{-\frac{\kappa_0}{20}|{\psi}_n - {\psi}^m_0|}e^{\frac{-j2\pi|{\psi}_n - {\psi}^m_0|}{\lambda_g}},
    \end{aligned}
\end{equation} 
where $\lambda_g=\frac{\lambda}{n_{\text{eff}}}$ is the guided wavelength, $n_{{\text{eff}}}$ is the effective refractive index of the dielectric waveguide, and $\kappa_0$ is the average attenuation factor along the dielectric waveguide in dB/m \cite{yeh2008kappa}.
The signal received by the PA might reradiate into free space from another PA along the same waveguide, which makes the signal model mathematically intractable. Thus, we suppose that each segment has only one activated PA, that is, $N_m = 1$ for $\forall m \in \mathcal{M}$. The uplink channel coefficient can be given by
\begin{equation}
    \label{uplinkchannel}
    H_k(\bm{\psi}^m_n, \bm{\psi}^m_0, \bm{\psi}_k) = h_i(\bm{\psi}^m_n, \bm{\psi}^m_0) h_o(\bm{\psi}_k, \bm{\psi}^m_n).
\end{equation}

The performance of segmented waveguide-enabled JCC-PASS framework is affected by the connection mechanism between the feed point and the RF front-end at the BS. The three basic operating protocols are utilized \cite{ouyang2025uplink}, i.e., segment selection (SS), segment aggregation (SA), and segment multiplexing (SM).

\subsubsection{Segment Selection} Only one segment of the waveguide is selected to connect the RF chain. The simple switching mechanism is adopted by this protocol for low complexity. The received signal via the SS-based uplink can be expressed as $y^{\text{SS}} = H_k(\bm{\psi}^{\overline{m}}_n, \bm{\psi}^{\overline{m}}_0, \bm{\psi}_k)x_k + n_{\overline{m}}$,
where $\overline{m}$ is the selected index of the segment, $n_{\overline{m}}\sim \mathcal{CN}(0,\sigma^2)$ follows additive white Gaussian noise (AWGN) with variance $\sigma^2$, $x_k$ is the transmit signal from the UE $k$.

\subsubsection{Segment Aggregation} All segments are aggregated, and then connected to the RF chain for baseband processing. Thus, the aggregated signal at the BS is expressed as $y^{\text{SA}} = \sum_{m=1}^{M} H_k(\bm{\psi}^m_n, \bm{\psi}^m_0, \bm{\psi}_k)(x_k + n_m)$,
where $\sum_{m=1}^{M} n_m \sim \mathcal{CN}(0, M\sigma^2)$ denotes the aggregated uplink noise.

\subsubsection{Segment Multiplexing} Each segment is connected to its own specialized RF chain. We stack the received signal of each segment into the received signal vector as $\mathbf{y}^{\text{SM}} = [y_1,y_2,\dots,y_M]^T = \mathbf{H}_k x_k + \mathbf{n}$,
where $\mathbf{H}_k = [H_k]^{M}_{m=1} \in \mathbb{C}^{M \times 1}$, and $\mathbf{n} = [n_1,n_2,\dots,n_M]^T \sim \mathcal{CN}(\mathbf{0},\sigma^2\mathbf{I}_M)$ denotes the noise vector.

\subsection{Signal Model}

\subsubsection{Computation}
In the proposed JCC-PASS framework, each UE performs sensing and computation simultaneously and transmit a superposition-coded signal towards BS. The computation symbol $d_{k}$ and the sensing symbol $s^{\prime}_{k}$ are measured as the values of parameters at the UE $k$, respectively.
Specifically, we assume that the UE records the one parameter data steam to compute and senses one data stream from environment and humans, which generates the corresponding scalars $d_k$ and $s^{\prime}_k$ for communication and computation. Following the nomographic function model \cite{gold2015nomog,qi2021iscc5g}, the BS can compute the nomographic function as $q_c(d_k)=f_c(\sum_{k=1}^{K}I_k(d_k))$,
\begin{equation}
    \label{q_c}
    q_c(d_k)=f_c(\sum_{k=1}^{K}I_k(d_k)),
\end{equation}
where $f_c(\cdot)$ and $I_{k}$ are the post-processing function at the BS and the pre-processing functions at the UE $k$, respectively. The UE $k$ computes the corresponding transmitted computation signal after pre-processing at UE $k$, denoted by $s_k = I_k(d_k)$, and transmits the superposition coded transmit signal,
\begin{equation}
    \label{x_k}
    x_k = w_k s_k + v_k s^{\prime}_{k},
\end{equation}
where $w_k$ denotes the transmit computation beamforming, and $v_k$ represents the transmit sensing beamforming for the sensing signal $s^{\prime}_k$. Based on the above segmented waveguide-enabled PASS, the general expression of the signal received at the feed point of segment $m$ can be given by \eqref{feed_signal},
\begin{equation}
    \label{feed_signal}
    \begin{aligned}
        & y_m = \sum_{k=1}^{K}H_k(\bm{\psi}^m_n, \bm{\psi}^m_0, \bm{\psi}_k) (w_k s_k + v_k s^{\prime}_k) + n_m, \\
        & = \!\sum_{k=1}^{K}\!H_k(\!\bm{\psi}^m_n\!,\! \bm{\psi}^m_0\!, \!\bm{\psi}_k\!)  w_k s_k\!+\!\sum_{k=1}^K\! H_k(\bm{\psi}^m_n\!,\! \bm{\psi}^m_0\!,\! \bm{\psi}_k\!) v_k s^{\prime}_k +\!n_m,
    \end{aligned}
\end{equation}
where $n_m \sim \mathcal{CN}(0,\sigma^2)$ follows additive white Gaussian noise (AWGN) with variance $\sigma^2$.
Let $s=\sum_{k=1}^{K}s_k = \sum_{k=1}^{K}I_k(d_k)$ be the targeted signal at the BS. The general computation distortion of the proposed JCC-PASS is measured by the MSE \cite{qi2021iscc5g,cheng2014mse},
\begin{equation}
    \label{MSE_function}
    \begin{aligned}
        & \text{MSE}(z,w_k,v_k) = \mathbb{E}[(x^{\prime}-s)(x^{\prime}-s)^H],\\
        & = \left|\sum_k z^H H_k w_k-1\right|^2 + \sum_k|z^H H_k v_k|^2 + |z|^2 \sigma^2,
    \end{aligned}
\end{equation}
where the first term $\left|\sum_k z^H H_k w_k-1\right|^2$ means the mismatching of aggregation, the second term $\sum_k|z^H H_k v_k|^2$ is the sensing leakage, and the third term $|z|^2 \sigma^2$ is the noise.

\subsubsection{Uplink PASS Communication}
Based on the three protocols, the received signal can be rewritten:

i) For SS, the received signal is rewritten by
\begin{equation}
    \label{ss_receivedsignal}
    \begin{aligned}
        y^{\text{SS}} = & \sum_{k=1}^{K} H_k(\bm{\psi}^{\overline{m}}_n, \bm{\psi}^{\overline{m}}_0, \bm{\psi}_k) w_k s_k +\\
        & \sum_{k=1}^{K} H_k(\bm{\psi}^{\overline{m}}_n, \bm{\psi}^{\overline{m}}_0, \bm{\psi}_k)  v_k s^{\prime}_k + n_{\overline{m}}.
    \end{aligned}
\end{equation}

ii) As for SA, the received signal can be rewritten as
\begin{equation}
    \label{sa_receivedsignal}
    \begin{aligned}
    y^{\text{SA}} = \sum_{k=1}^{K}\sum_{m=1}^{M}H_k(\bm{\psi}^m_n, \bm{\psi}^m_0, \bm{\psi}_k) [w_k s_k + v_k s^{\prime}_k] + \sum_{m=1}^{M} n_m.
    \end{aligned}
\end{equation}

iii) For SM, the received signal is updated by
\begin{equation}
    \label{sm_receivedsignal}
    \mathbf{y}^{\text{SM}} = \sum_{k=1}^{K} \mathbf{H}_k w_k s_k +  \sum_{k=1}^{K} \mathbf{H}_k  v_k s^{\prime}_k + \mathbf{n}.
\end{equation}

For simplicity, we assume $\mathbb{E}[|s^{\prime}_{k}|^2] = 1$. Let $z \in \mathbb{C}$ denote the received computation beamforming at the BS. The general expression of the recovered computation signal at the BS can be written as
\begin{equation}
    \label{x'}
    \begin{aligned}
        x^{\prime} = & z^H\sum_{k=1}^{K} H_k(\bm{\psi}^m_n, \bm{\psi}^m_0, \bm{\psi}_k) w_k s_k \\
        & + z^H\sum_{k=1}^{K} H_k(\bm{\psi}^m_n, \bm{\psi}^m_0, \bm{\psi}_k) v_k s^{\prime}_k + z^H n_m,
    \end{aligned}
\end{equation}
where $z$ denotes the receive computation beam.
Moreover, the general expression of the recovered sensing signal at the BS from UE $k$ can be formulated as
\begin{equation}
    \label{receivedsensing}
    \begin{aligned}
        y^{\prime} = & u^H_k H_k(\bm{\psi}^m_n,\! \bm{\psi}^m_0,\! \bm{\psi}_k) v_k s^{\prime}_k + u^H_k\! \sum_{i\neq k}^{K}\! H_k(\bm{\psi}^m_n,\! \bm{\psi}^m_0,\! \bm{\psi}_k) v_i s^{\prime}_i \\
        & + u^H_k \sum_{i=1}^{K} H_k(\bm{\psi}^m_n, \bm{\psi}^m_0, \bm{\psi}_k) w_k s_i + u^H_k n_m,
    \end{aligned}
\end{equation}
where $u_k$ is the receive sensing beam at the BS from UE $k$.
Further, the corresponding the signal-to-interference-plus-noise ratio (SINR) can be expressed as
\begin{equation}
    \label{SINR_k}
        \text{SINR}_k = \frac{|u_k H_k v_k|^2}{\sum_{i\neq k}^{K}|u_k H_i v_i|^2 \! + \! \sum_{i=1}^{K}\! |u_k H_i w_i|^2 \! + \! \sigma^2|u_k|^2}.
\end{equation}

Under three different protocols, we summarize the corresponding expressions of their MSE and SINR in Table \ref{tab:mse_wsr}.

\begin{table*}[htbp]
    \centering
    \caption{MSE and SINR Expressions Under Three Uplink Protocols}
    \label{tab:mse_wsr}
    \renewcommand{\arraystretch}{1.5}
    \setlength{\tabcolsep}{3.5pt} 
    \scriptsize
    \begin{tabular}{ccc} %
    \toprule
    \textbf{Protocol} & \textbf{MSE} & \textbf{SINR}\\
    \midrule
    \textbf{SS} & $\text{MSE}^{\text{SS}} =$ \eqref{MSE_function} & $\text{SINR}^{\text{SS}}_k =$ \eqref{SINR_k}\\
    \textbf{SA} & $\text{MSE}^{\text{SA}} =\left|z^H\!\sum_{k}\sum_{m}\! H_k w_k\!-\!1\right|^{2}\!+\!|z|^{2}\!\sum_{k}\!\left|\sum_{m}\!H_k v_k\right|^{2}\!+\!|z|^{2}\!M\sigma^{2}$ & $\text{SINR}^{\text{SA}}_k=\frac{|\sum_{m}H_k v_k|^{2}}{\sum_{i\neq k}|\sum_{m} H_i v_i|^{2}+\sum_{i}|\sum_{m} H_i w_i|^{2}+M\sigma^{2}}$\\
    \textbf{SM} & $\text{MSE}^{\text{SM}}=\|\mathbf{z}^H\!\sum_{k}\!\mathbf{H}_k w_k\!-\!1\|_2^{2}\!+\!\|\mathbf{z}\|^{2}\!\sum_{k}\!\|\mathbf{H}_k v_k\|_2^{2}\!+\!\|\mathbf{z}\|^{2}\sigma^{2}\|\mathbf{n}\|_{2}^{2}$ & $\text{SINR}_k^{\text{SM}}=\frac{\|\mathbf{H}_k v_k\|_2^{2}}{\sum_{i\neq k}\|\mathbf{H}_i v_i\|_2^{2}+\sum_{i}\|\mathbf{H}_i w_i\|_2^{2}+\sigma^{2}\|\mathbf{n}\|_{2}^{2}}$\\
    \bottomrule
    \end{tabular}
    \end{table*}

\section{MSE Minimization Design for JCC-PASS}
In this section, we design the computation-oriented MSE minimization problem for JCC-PASS framework under three protocols, i.e., SS, SA, and SM. Further, we aim to solve the corresponding problems by AO-MMSE-based algorithm under these three protocols in the following three subsections. Moreover, we compare the proposed PASS framework with the conventional PASS in the average in-waveguide gain, and we mathematically prove that the proposed JCC-PASS outperforms the conventional PASS. 

\subsection{Problem Formulations}
In the proposed JCC-PASS framework, we aim to optimize the MSE for computation performance. It can be observed that the transmit beamforming $w_k$ and $v_k$ at the UE $k$ and receive beams $z$ and $u_k$ at the BS all affect the computation and communication performances. Therefore, it is necessary to design the computation-oriented MSE minimization problem to optimize the transmit beamforming at the UEs and receive beams at the BS. The general computation-oriented problem can be formulated as
\begin{subequations}
    \label{P_0}
    \begin{align}
        \mathbf{P}_0: \quad & \min_{z, w_k, v_k, u_k}\text{MSE}\left(z,w_k,v_k\right)
        \label{min_MSE0} \\
        \mathrm{s.t.} \quad & \log_{2}\left(1+ \text{SINR}_k \right)\geq r^{\min}_k, \forall k \in \mathcal{K}
        \label{rateconstraint0}\\
        & |{\psi}^{m}_{n'} - {\psi}^{m}_n| \ge \Delta \ge 0, \forall n \ne n', n, n' \in \mathcal{N},
        \label{PAconstraint0}\\
        & |w_k|^{2} + |v_k|^{2} \leq P^{\max}_k, \forall k \in \mathcal{K},
        \label{powerconstraint0}
       \end{align}
\end{subequations}
where \eqref{rateconstraint0} ensures the minimum data rate of each UE, \eqref{PAconstraint0} guarantees the minimum inter-antenna spacing required to mitigate mutual coupling effects, and \eqref{powerconstraint0} guarantees the maximum power budget of each UE for transmission, $r^{\min}_k$ is the minimum data rate of the UE $k$ and $P^{\max}_k$ is the maximum transmit power of the UE $k$. Since the strong coupling among transmit beamforming and receive beams, this problem \eqref{min_MSE0} is NP-hard and nonconvex. The three protocols' MSE expressions are summarized in the Table \ref{tab:mse_wsr}.

The MSE expressions of three protocols contain multiplicative and quadratic terms of $\{z, w_k, v_k\}$, while the SINR constraint introduces a fractional coupling between the transmit beamforming set $\{w_k, v_k\}$ and the receive beam set $\{z, u_k\}$. Hence, the computation-oriented MSE minimization problem is nonconvex, we separate it into two subproblems based on AO. Specifically, one is the receiver-side subproblem that is convex for the fixed transmit beamforming set and yields the MMSE receiver; another one is the transmitter-side subproblem that is strictly convex quadratic function for the fixed receive beam set.
Mathematically, the optimization variables, objective function structures and constraint forms are identical for the three protocols. Therefore, the AO and MMSE-based solution process are common to all three protocols.

\subsection{Solution of MSE Minimization for SS}
Based on \eqref{MSE_function} and Table \ref{tab:mse_wsr}, with the fixed transmit beamforming set $(w_k, v_k)$ the receiver-side subproblem can be formulated as
\begin{subequations}
    \label{P_0.0a}
    \begin{align}
        \mathbf{P}_{0.0.\text{a}}: \quad & \min_{z, u_k}\text{MSE}^{\text{SS}}
        \label{min_MSE_ss0} \\
        \mathrm{s.t.} \quad & \eqref{rateconstraint0}-\eqref{powerconstraint0},
       \end{align}
\end{subequations}
then, the MMSE-based receivers can be derived as
\begin{equation}
    \label{z_mse_ss}
    z^{\text{M-SS}} = \frac{\sum_{k} H_k w_k}{|\sum_{k}H_k w_k|^2 + \sum_{k}|H_k v_k|^2 + \sigma^2},
\end{equation} 
and
\begin{equation}
    \label{u_MSE_SS}
    u_k^{\text{M-SS}} =\frac{H_k v_k}{|\sum_{i}^{K}H_i w_i|^2 + \sum_{i\neq k}|H_i v_i|^2 + \sigma^2}.
\end{equation}

Further, another transmitter-side MSE subproblem under SS protocol can be formulated as
\begin{subequations}
    \label{P_0.0b}
    \begin{align}
        \mathbf{P}_{0.0.\text{b}}: \quad & \min_{w_k, v_k}|z^{\text{M-SS}}\!\sum_{k} \!H_k w_k\! -\! 1|^2\! + \!\sum_{k}|z|^2|H_k|^2|v_k|^2,
        \label{min_MSE_ss1} \\
        \mathrm{s.t.} \quad & \eqref{rateconstraint0}-\eqref{powerconstraint0},
       \end{align}
\end{subequations}
where this objective is convex quadratic and linear for $w_k$ and $|v_k|^2$, respectively. Substitute \eqref{rateconstraint0}, we can obtain
\begin{equation}
    \label{rate_ss_inequality}
    \frac{1}{\gamma^{\text{SS}}_k}|u^\text{M-SS}_k H_k v_k|^2 \! \ge \!\sum_{i=1}^{K}\!|u_k H_i w_i|^2\! +\! \sum_{i\neq k}^{K}\!|u_k H_i v_i|^2\! + \!\sigma^2|u_k|^2,
\end{equation}
where the term $\sum_{i=1}^{K}|u_k H_i w_i|^2$ represents the computation leakage, $\sum_{i\neq k}^{K}|u_k H_i v_i|^2$ means the sensing interference, and $\gamma^{\text{SS}}_k = 2^{r^{\min}_k}-1$. To obtain the convexity, we employ the SDR method to convert \eqref{P_0.0b} into
\begin{subequations}
    \label{P_0.0c}
    \begin{align}
        \mathbf{P}_{0.0.\text{c}}: & \min_{w_k, |v_k|^2}|z^{\text{M-SS}}\sum_{k} H_k w_k \! - \!1|^2\! + \!|z|^2\sum_{k}|H_k|^2|v_k|^2,
        \label{min_MSE_ss2} \\
        \mathrm{s.t.} \quad & \eqref{rate_ss_inequality}, \eqref{rateconstraint0}-\eqref{powerconstraint0},
       \end{align}
\end{subequations}
where this transformed problem is convex and can be solved by the interior point method. Without power constraint, the optimal transmit beamforming $\tilde{w}_k$ with the fixed receive beam set can be derived as $\tilde{w}_k = \frac{z^{\text{M-SS}}H_k}{\sum_{i=1}^{K}|z|^2|H_i|^2}$. Next, we consider the power and rate constraints to make $\tilde{w}_k$ as the scale contraction to maintain the optimal direction, which can be shrunk inward uniformly to the tightest constraint with the contraction $\rho^{\text{M-SS}}= \sqrt{\frac{P^{\max}_k-|v_k|^2}{\sum_{i}^K|\tilde{w}_i|^2}}$,
\begin{equation}
    \label{w_mse_ss}
    w^{\text{M-SS}}_k=\rho^{\text{M-SS}}\tilde{w}_k = \sqrt{\frac{P^{\max}_k-|v_k|^2}{\sum_{i}^K|\tilde{w}_i|^2}}\cdot \frac{z^{\text{M-SS}}H_k}{\sum_{i=1}^{K}|z|^2|H_i|^2}.
\end{equation}
Finally, the closed-form solution of $v_k$ is given by
\begin{equation}
    \label{v_mse_ss}
    v_k^{\text{M-SS}} = \sqrt{|v_k|^2}e^{j\angle (H_k u^{\text{M-SS}}_k)}.
\end{equation}

For the SS protocol, only one segment and one PA are connected to the RF chain. According to the channel model in \eqref{uplinkchannel}, the average in-waveguide propagation gain for SS can be written as
\begin{equation}
    \label{ss_gain}
    \begin{aligned}
        A^{\text{SS}} & \triangleq \frac{1}{L}\int_{0}^{L}10^{-\frac{\kappa_0}{10}x}\mathrm{d}x =\frac{1}{L}\frac{1-10^{-\frac{\kappa_0}{10}L}}{\frac{\kappa_0}{10}\ln10} \\
        & =\frac{1-\mathrm{e}^{-2\alpha L}}{2\alpha L}=\frac{1-\mathrm{e}^{-2\alpha\frac{D^{\text{x}}_0}{M}}}{2\alpha\frac{D_0^{\text{x}}}{M}},
    \end{aligned}
\end{equation}
where $\alpha = \frac{\kappa_0\ln(10)}{20}$. Unlike the proposed segment waveguide-enabled JCC-PASS framework, the average in-waveguide propagation gain of the conventional PASS framework is
\begin{equation}
    \label{conven_gain}
    A^{\text{C}} \triangleq \frac{1}{D_0^{\text{x}}}\int_{0}^{D_0^{\text{x}}}10^{-\frac{\kappa_0}{10}x}\mathrm{d}x = \frac{1-\mathrm{e}^{-2\alpha D_0^{\text{x}}}}{2\alpha D_0^{\text{x}}}.
\end{equation}

\begin{lemma}
    \label{optimal_PA_SS}
    The PA activated within the segment is the one closest to the UE, where the segment index is $\overline{m}^{\star} = \left\lceil\frac{\psi^{\text{x}}_k - \psi^{1}_{0}}{L}\right\rceil$. Moreover, the closed-form activated position of PA is derived as
    \begin{equation}
        \label{PA_position}
        \psi_n^{\overline{m}^\star}=\left\{
            \begin{array}
                {ll}\psi_n^{\overline{m}^\star} & d_{0}\geq\frac{1-(2\alpha |\psi^{\text{x}}_k - \psi_n^{{\overline{m}^{\star}}}|-1)^2}{4\alpha^2} \\
                \psi^{\text{x}}_k+\frac{-1+\sqrt{1-4\alpha^2 d_0}}{2\alpha} & \mathrm{else}
            \end{array}\right.,
    \end{equation}
    where $d_{0} = [(\psi^{\text{y}}_k - D^{\text{y}}_0)^2 + (D^{\text{z}}_0)^2]$. Notably, only in the case that $n=1, k=K=1$, there is the optimal PA position.
\end{lemma}

\begin{IEEEproof}
    Mathematically, $e^x > x+1$ for $x>0$, we can get $e^{\frac{2\alpha D^{\text{x}}_0}{M}} > \frac{2\alpha D^{\text{x}}_0}{M} + 1$ and $(\frac{2\alpha D^{\text{x}}_0}{M}+1)e^{-\frac{2\alpha D^{\text{x}}_0}{M}} < 1$. In addition, from \eqref{ss_gain}, the first-order derivative of the in-waveguide propagation gain with respect to $M$ is given by
    \begin{equation}
        \label{first_derivative_gain}
        \tfrac{\partial A^{\text{SS}}}{\partial M} = \frac{1}{2\alpha D^{\text{x}}_0}\left(1-\left(1+\frac{2\alpha D^{\text{x}}_0}{M}\right)e^{-\frac{2\alpha D^{\text{x}}_0}{M}}\right) > 0.
    \end{equation}

    The second-order derivative of $A^{\text{SS}}$ is written as
    \begin{equation}
        \label{second_derivative_gain}
        \tfrac{\partial^2 A^{\text{SS}}}{\partial M^2} = -\frac{2\alpha D^{\text{x}}_0}{M^3} e^{\frac{2\alpha D^{\text{x}}_0}{M}} < 0,
    \end{equation}
    the $M$ exists to maximize the in-waveguide propagation gain.
    From the \eqref{SINR_k}, the activated PA position is closest to the UE $k$ in the segment $\overline{m}^{\star}$, where $\overline{m}^{\star} = \left\lceil\frac{\psi^{\text{x}}_k - \psi^{1}_{0}}{L}\right\rceil$ can be derived. Substituting the expression of $\overline{m}^{\star}$ into \eqref{SINR_k} and \eqref{ss_gain}, we can obtain the closed-form PA position \eqref{PA_position}.

    The closed-form PA position is jointly determined by the path-loss attenuation along the waveguide and the free-space distance between the PA and the UE \cite{ouyang2025array}. Specifically, when the UE is located farther away from the waveguide, i.e., $d_0$, the PA activation point should move slightly toward the segment origin to balance the attenuation of the waveguide propagation and the free-space path loss. Conversely, when the UE is near the waveguide, the PA position tends to coincide with the user's projection on the $x$-axis. Therefore, aligning the PA position with the UE projection minimizes the composite path-loss component and achieves the maximum gain for the SS protocol.
\end{IEEEproof}

\begin{remark}
    \label{inwave_loss_super_pass}
    The segmented PASS always outperforms conventional PASS with a single continuous waveguide lower in in-waveguide propagation loss. Further, this advantage becomes more pronounced with either a larger service region (i.e., larger $D^{\text{x}}_0$) or a higher number of segments $M$.
\end{remark}

\begin{IEEEproof}
    The fractional comparison with average in-waveguide propagation gain of JCC-PASS and conventional PASS can be formulated as
    \begin{equation}
        \label{gain_compare}
        \frac{A^{\text{SS}}}{A^{\text{C}}}= \frac{(1-\mathrm{e}^{-2\alpha\frac{D^{\text{x}}_0}{M}})M}{1-\mathrm{e}^{-2\alpha D_0^{\text{x}}}},
    \end{equation}
    where the fraction is equal to 1 as $M=1$, and is monotonically increasing with $M$. Moreover, it is also monotonically increasing with $D_0^{\text{x}}$. That means the JCC-PASS framework has superiority over conventional PASS of the average in-waveguide propagation gain, which is more significant as segment or waveguide length increases.
\end{IEEEproof}

\subsection{Solution of MSE Minimization for SA}
Based on the Table \ref{tab:mse_wsr} and the above solution procedure of MSE minimization problem, this problem is equally nonconvex as that of the SS protocol, we employ the same process to solve this problem under the SA protocol. The first receiver-side MSE subproblem can be formulated as
\begin{subequations}
    \label{P_0.1a}
    \begin{align}
        \mathbf{P}_{0.1.\text{a}}: \quad & \min_{z, u_k}\text{MSE}^{\text{SA}}
        \label{min_MSE_sa0} \\
        \mathrm{s.t.} \quad & \eqref{rateconstraint0}-\eqref{powerconstraint0},
       \end{align}
\end{subequations}
the corresponding MMSE-form receivers can be derived as
\begin{equation}
    \label{z_mse_sa}
    z^{\text{M-SA}} = \frac{\sum^{K}_{k} \sum_{m}^{M} H_k w_k}{|\sum^{K}_{k} \sum_{m}^{M} H_k w_k|^2 + \sum^{K}_{k}|\sum^{M}_{m} H_k v_k|^2 + M\sigma^2},
\end{equation} 
and
\begin{equation}
    \label{u_MSE_SA}
    u_k^{\text{M-SA}} =\frac{\sum_{m}^{M} H_k v_k}{|\sum_{i}^{K} \sum_{m}^{M} H_i w_i|^2 + \sum_{i\neq k}^{K}|\sum_{m}^{M} H_i v_i|^2 + M\sigma^2}.
\end{equation}

The other transmitter-side subproblem can be written as
\begin{subequations}
    \label{P_0.1b}
    \begin{align}
        \mathbf{P}_{0.1.\text{b}}: & \min_{w_k, v_k}|z^{\text{M-SA}}\!\sum_{k} \sum_{m}\!H_k w_k\! -\! 1|^2\! + \!\sum_{k}\!|z|^2|\sum_{m}H_k|^2|v_k|^2,
        \label{min_MSE_sa1} \\
        \mathrm{s.t.} \quad & \eqref{rateconstraint0}-\eqref{powerconstraint0},
       \end{align}
\end{subequations}
similarly, the objective is also convex quadratic and linear for $w_k$ and $|v_k|^2$, respectively. The rate constraint can be transformed into
\begin{equation}
    \label{rate_sa_inequality}
    \begin{aligned}
        \frac{1}{\gamma^{\text{SA}}_k}|u^\text{M-SA}_k \sum_{m}H_k v_k|^2 \ge & \sum_{i=1}^{K}|u_k \sum_{m}H_i w_i|^2 + M\sigma^2|u_k|^2 \\
        & + \sum_{i\neq k}^{K}|u_k \sum_{m}H_i v_i|^2,
    \end{aligned}
\end{equation}
where $\gamma^{\text{SA}}_k = 2^{r^{\min}_k}-1$. To obtain the convexity, the SDR method is employed to convert \eqref{P_0.1b} into
\begin{subequations}
    \label{P_0.1c}
    \begin{align}
        \mathbf{P}_{0.1.\text{c}}: & \min_{w_k, |v_k|^2}|z^{\text{M-SA}}\!\sum_{k} \sum_{m} H_k w_k \! - \!1|^2\! + \!|\!z\!|^2\sum_{k}\!|\!\sum_{m}\!H_k\!|^2\!|\!v_k\!|^2,
        \label{min_MSE_sa2} \\
        \mathrm{s.t.} \quad & \eqref{rate_sa_inequality}, \eqref{rateconstraint0}-\eqref{powerconstraint0}.
       \end{align}
\end{subequations}

It can be seen that this problem is convex as \eqref{P_0.0c} and can be solved by the interior point method. Following the same process under the SS protocol, the closed-form solution of transmit beamforming set can be derived as
\begin{equation}
    \label{w_mse_sa}
    w^{\text{M-SA}}_k=\rho^{\text{M-SA}}\tilde{w}_k = \sqrt{\frac{P^{\max}_k-|v_k|^2}{\sum_{i}^K|\tilde{w}_i|^2}}\cdot \frac{z^{\text{M-SA}}\sum_{m} H_k}{\sum_{i=1}^{K}|z|^2|\sum_{m} H_i|^2},
\end{equation}
and
\begin{equation}
    \label{v_mse_sa}
    v_k^{\text{M-SA}} = \sqrt{|v_k|^2}e^{j\angle (\sum_{m} H_k u^{\text{M-SA}}_k)},
\end{equation}
where $\rho^{\text{M-SA}}= \sqrt{\frac{P^{\max}_k-|v_k|^2}{\sum_{i}^K|\tilde{w}_i|^2}}$. 

\subsection{Solution of MSE Minimization for SM}
The MSE minimization problem for SM is nonconvex, and hence, we also adopt the AO method to deal with it. Different from the SS and SA protocols, the receive beams $\mathbf{z} \in \mathbb{C}^{M \times 1}, \mathbf{u}_k \in \mathbb{C}^{1\times M}$ are vectors. From \eqref{MSE_function} and Table \ref{tab:mse_wsr}, with the fixed transmit beamforming set $w_k, v_k \in \mathbb{C}$ the receiver-side subproblem can be formulated as
\begin{subequations}
    \label{P_0.2a}
    \begin{align}
        \mathbf{P}_{0.2.\text{a}}: \quad & \min_{\mathbf{z}, \mathbf{u}_k}\text{MSE}^{\text{SM}}
        \label{min_MSE_sm0} \\
        \mathrm{s.t.} \quad & \eqref{rateconstraint0}-\eqref{powerconstraint0},
       \end{align}
\end{subequations}
additionally, the MMSE-form receivers can be derived as
\begin{equation}
    \label{z_mse_sm}
    \mathbf{z}^{\text{M-SM}} = \frac{\sum_{k} \mathbf{H}_k w_k}{(\sum_{k}\mathbf{H}_k w_k)(\sum_{j}\mathbf{H}_j w_j)^H + \sum_{k}|v_k|^2\mathbf{H}_k \mathbf{H}^H_k + \sigma^2},
\end{equation} 
the receive sensing beam at the BS can be formulated as
\begin{equation}
    \label{u_MSE_SM}
    \mathbf{u}_k^{\text{M-SM}} =\frac{\mathbf{H}_k v_k}{\sum_{i}^{K}|w_i|^2\mathbf{H}_i\mathbf{H}^H_i + \sum_{i\neq k}|v_i|^2\mathbf{H}_i\mathbf{H}^H_i + \sigma^2}.
\end{equation}

Further, the transmitter-side subproblem is expressed as
\begin{subequations}
    \label{P_0.2b}
    \begin{align}
        \mathbf{P}_{0.2.\text{b}}: & \min_{w_k, v_k}|\mathbf{z}^H\!\sum_{k} \!\mathbf{H}_k w_k\! -\! 1|^2\! + \!\sum_{k}\!\mathbf{z}^H \mathbf{H}_k |v_k|^2\mathbf{z},
        \label{min_MSE_sm1} \\
        \mathrm{s.t.} \quad & \eqref{rateconstraint0}-\eqref{powerconstraint0},
       \end{align}
\end{subequations}
where this objective is convex quadratic and linear for $w_k$ and $|v_k|^2$, respectively. The rate constraint is converted into
\begin{equation}
    \label{rate_sm_inequality}
        \frac{1}{\gamma^{\text{SM}}_k}|\mathbf{u}^\text{M-SM}_k \mathbf{H}_k v_k|^2 \ge \sum_{i=1}^{K}|\mathbf{u}_k \mathbf{H}_i w_i|^2 + \sigma^2|\mathbf{u}_k|^2 + \sum_{i\neq k}^{K}|\mathbf{u}_k \mathbf{H}_i v_i|^2,
\end{equation}
where $\gamma^{\text{SM}}_k = 2^{r^{\min}_k}-1$. We utilize the SDR method for the convexity, by converting \eqref{P_0.2b} into
\begin{subequations}
    \label{P_0.2c}
    \begin{align}
        \mathbf{P}_{0.2.\text{c}}: & \min_{w_k, |v_k|^2}|\mathbf{z}^{\text{M-SM}}\!\sum_{k} \mathbf{H}_k w_k \! - \!1|^2\! + \!|z|^2\sum_{k}|\mathbf{H}_k|^2|v_k|^2,
        \label{min_MSE_sm2} \\
        \mathrm{s.t.} \quad & \eqref{rate_sm_inequality}, \eqref{rateconstraint0}-\eqref{powerconstraint0}.
       \end{align}
\end{subequations}

This problem is convex and can be addressed by the interior point method. Thus, the closed-form solution of transmit beamforming set can be derived by
\begin{equation}
    \label{w_mse_sm}
    w^{\text{M-SM}}_k=\rho^{\text{M-SM}}\tilde{w}_k = \sqrt{\frac{P^{\max}_k-|v_k|^2}{\sum_{i}^K|\tilde{w}_i|^2}}\cdot \frac{\mathbf{z}^{\text{M-SM}} \mathbf{H}_k}{\sum_{i=1}^{K}|\mathbf{z}|^2| \mathbf{H}_i|^2},
\end{equation}
and
\begin{equation}
    \label{v_mse_sm}
    v_k^{\text{M-SM}} = \sqrt{|v_k|^2}e^{j\angle (\mathbf{H}_k \mathbf{u}^{\text{M-SM}}_k)},
\end{equation}
where $\rho^{\text{M-SM}}= \sqrt{\frac{P^{\max}_k-|v_k|^2}{\sum_{i}^K|\tilde{w}_i|^2}}$, and satisfies $|v_k^{\text{M-SM}}|^2 \in [P^{\max}_k-|w_k|^2]$.

\begin{algorithm}[htbp]
    \footnotesize
    \caption{The AO-MMSE-Based MSE Minimization Algorithm for SS, SA and SM Protocols}
    \label{alg:ao_mmse}
    \begin{algorithmic}[1]
    \REQUIRE Protocol $\mathsf{P}\in\{\text{SS},\text{SA},\text{SM}\}$:$H_k$ (SS), $\tilde H_k$ (SA), $\mathbf{H}_k$ (SM), $\sigma^2$; $M$ (SA/SM), $r_k^{\min}$ with $\gamma_k=2^{r_k^{\min}}-1$, $P_k^{\max}$.
    \ENSURE $\{w_k,v_k\}$, $z$ (SS/SA), $\mathbf{z}$, $\{\mathbf{u}_k\}$ for SM.
    \STATE Initialize a feasible $\{w_k^{(0)},v_k^{(0)}\}$ with $|w_k^{(0)}|^2+|v_k^{(0)}|^2\le P_k^{\max}$, $V_k^{(0)}=|v_k^{(0)}|^2$, $t = 0$.
    \REPEAT
      \STATE \textbf{Receiver-side subproblem:}
      \IF{$\mathsf{P}=\text{SS}$}
        \STATE Update $z^{(t)}$ and $u_k^{(t)}$ from \eqref{z_mse_ss} and \eqref{u_MSE_SS} with the fixed $\{w^{(t-1)}_k, v_k^{(t-1)}\}$.
      \ELSIF{$\mathsf{P}=\text{SA}$}
        \STATE Update $z^{(t)}$ and $u_k^{(t)}$ from \eqref{z_mse_sa} and \eqref{u_MSE_SA} with the fixed $\{w^{(t-1)}_k, v_k^{(t-1)}\}$.
      \ELSE
        \STATE Update $\mathbf{z}^{(t)}$ and $\mathbf{u}_k^{(t)}$ from \eqref{z_mse_sm} and \eqref{u_MSE_SM} with the fixed $\{w^{(t-1)}_k, v_k^{(t-1)}\}$.
      \ENDIF
      \STATE \textbf{Transmitter-side subproblem:}
      \IF{$\mathsf{P}=\text{SS}$}
        \STATE Update $w^{(t)}_k, v_k^{(t)}$ by solving \eqref{P_0.0c}.
      \ELSIF{$\mathsf{P}=\text{SA}$}
        \STATE Update $w^{(t)}_k, v_k^{(t)}$ by solving \eqref{P_0.1c}.
      \ELSE
        \STATE Update $w^{(t)}_k, v_k^{(t)}$ by solving \eqref{P_0.2c}.
      \ENDIF
      \STATE $t\gets t+1$
    \UNTIL{relative MSE change $\le \varepsilon$}
    \end{algorithmic}
    \end{algorithm}

\begin{lemma}
    \label{MMSE_receiver_proof}
    The closed-form solution of MSE minimization problem can be only obtained by MMSE receiver $a_k$ with the equality $a_k = \frac{1}{1+\text{SINR}_k}$. 
\end{lemma}
\begin{IEEEproof}
Let's prove this under the SM protocol, which can be reduced to similar scalar-form proof. From the Table. \ref{tab:mse_wsr}, we can rewrite the MSE expression by
\begin{equation}
    \label{MSE_SM}
    \begin{aligned}
        \text{MSE}^{\text{SM}} & =\sum_{i=1}^{K}\mathbf{u}_{k}^{H}\mathbf{H}_{i} v_i v_i^{H}\mathbf{H}_{i}^{H}\mathbf{u}_{k} +\sum_{i=1}^{K}\mathbf{u}_{k}^{H}\mathbf{H}_{i} w_{i} w_{i}^{H} \mathbf{H}_{i}^{H}\mathbf{u}_{k}  \\
        & \quad -\mathbf{u}_{k}^{H}\mathbf{H}_{k} v_k - v_k^{H}\mathbf{H}_{k}^{H}\mathbf{u}_{k} +\sigma^{2}\mathbf{u}^H_k \mathbf{u}_{k} + 1 \\
        & = \mathbf{u}_k^H\bm{\Omega}_k\mathbf{u}_k-\mathbf{u}_k^H\mathbf{H}_k v_k - v_k^H \mathbf{H}_k^H \mathbf{u}_k+1,
    \end{aligned}
\end{equation}
for simplicity, $\bm{\Omega}_k$ is denoted by 
\begin{equation}
    \label{Omega}
    \bm{\Omega}_k =\sum_{i=1}^{K}\mathbf{H}_{i}\left(v_i v_i^{H} + w_{i} w_i^{H} \right)\boldsymbol{H}_{i}^{H} +\sigma^{2}\mathbf{I}.
\end{equation}

Based on the above equations, we take the derivative of \eqref{MSE_SM} w.r.t. $\mathbf{u}_k$ and set it to zero gives
\begin{equation}
    \label{derivative_mse_omega}
    \frac{\partial\text{MSE}^{\text{SM}}}{\partial \mathbf{u}_{k}^{\text{M-SM}}}=\bm{\Omega}_{k}\mathbf{u}^{\text{M-SM}}_{k}-\mathbf{H}_{k} v_{k}=\bm{0},
\end{equation}
where it can be obtained by
\begin{equation}
    \label{u_mse_sm_proof}
    \mathbf{u}_k^{\text{M-SM}} = \bm{\Omega}_k^{-1}\mathbf{H}_k v_k,
\end{equation}
which adopts the MMSE receiver. Substituting \eqref{u_mse_sm_proof} into \eqref{MSE_SM}, the minimum MSE can be derived by
\begin{equation}
    \label{a_k}
    a_k = 1 - v_k^H \mathbf{H}^H_k \bm{\Omega}_k^{-H}\mathbf{H}_k v_k = \frac{1}{1+\text{SINR}_k^{\text{SM}}} = \text{MSE}^{\text{SS}}.
\end{equation}

It is proved that the minimum MSE and closed-form solution of MSE minimization problem can be obtained by maximizing the sensing SINR with the relationship of MMSE receiver $a_k = \frac{1}{1+\text{SINR}_k}$.
\end{IEEEproof}

In summary, the AO-MMSE-based algorithm for MSE minimization under SS, SA, and SM protocols is shown as Algorithm \ref{alg:ao_mmse}.

\subsection{Computational Complexity Analysis}

For the SS protocol, all variables are scalars involve only simple arithmetic operations. In each AO iteration, the MMSE receiver $z$ is updated by a closed-form scalar expression, which requires $O(K)$ additions and multiplications. The transmitter-side subproblem is solved by the linear system $\tfrac{1}{\gamma^{\text{SS}}_k}|H_k|^2 |v_k|^2 - \sum_{i\ne k}|H_i|^2 |v_i|^2 = \sum_i |H_i|^2|w_i|^2 + \sigma^2$.
When the special structure of the coefficient matrix is exploited, the system can be solved in $O(K)$ operations. Hence, the complexity of each iteration is $O(K)$ and the total computational complexity of the proposed algorithm under the SS protocol is $O(TK)$, where $T$ is the number of iterations.

The computational procedure of the SA protocol is identical to that of SS except that the aggregated channel $\sum^{M}_{m} H_k$ and the equivalent noise variance $M\sigma^2$ are employed in the receiver update. Therefore, the computation process remains a scalar MMSE operation with complexity $O(K)$. The transmitter-side subproblem follow the same expressions as in SS, and the structured linear system can again be efficiently solved in $O(K)$ time. Consequently, the computational complexity of each iteration is $O(K)$, and the overall computational complexity of the proposed algorithm under the SA protocol is $O(TK)$.

For the SM protocol, the receiver update involves an $M\times M$ matrix inversion $\mathbf{z}$ from \eqref{z_mse_sm}. The covariance matrix construction requires $O(M^2K)$ operations. If the optional sensing combiners $\{\mathbf{u}_k\}$ are updated by the MMSE rule, the same decomposition can be reused, resulting in an additional $O(KM^2)$ cost. For the transmitter-side subproblem, the inner products $\mathbf{z}^H\mathbf{h}_k$ and the scaling factor $\rho$ are computed in $O(KM)$, and the structured linear system for $|v_k|^2$ can be solved in $O(K)$. Therefore, the computational complexity of each iteration is $O(M^3+KM^2)$, and the total computational complexity of the proposed algorithm scales as $O\!\big(T(M^3+KM^2)\big)$.

\section{WSR Maximization Design for JCC-PASS}
In this section, we consider the communication-oriented WSR maximization problem for JCC-PASS framework under SS, SA, and SM protocols. Then, in the following three sections, we develop the AO-WMMSE-based algorithm to obtain the closed-form solutions. 

\subsection{Problem Formulation}
The general communication-oriented WSR maximization problem can be formulated as
\begin{subequations}
    \label{P_1}
    \begin{align}
        \mathbf{P}_1: \quad & \max_{z, w_k, v_k, u_k} \sum_{k=1}^{K}\theta_k \log_{2}\left(1+ \text{SINR}_k \right)
        \label{min_WSR1} \\
        \mathrm{s.t.} \quad & \text{MSE}\left(z,w_k,v_k\right) \le \zeta,
        \label{MSEconstraint1}\\
        & \eqref{PAconstraint0}-\eqref{powerconstraint0},
       \end{align}
\end{subequations}
where \eqref{MSEconstraint1} ensures the maximum tolerable computation error $\zeta$, $\theta_k >0$ denotes the weight coefficient of the sensing signal $s^{\prime}_k$ at the UE $k$. This problem \eqref{min_WSR1} is also NP-hard and nonconvex due to the same reason as that of the above computation-oriented problem.

The communication-oriented WSR maximization problems of these three protocols exhibit strong nonconvexity due to the logarithmic and multiplicative couplings among the transmit beamforming set $\{w_k, v_k\}$, the receive beam set $\{z, u_k\}$, and the auxiliary variables. In particular, the SINR terms introduce fractional structures between the transmit and receive domains, while the MSE constraints \eqref{MSEconstraint1} further complicate the feasible region. To tackle this challenge, this problem is efficiently decomposed into two subproblems by the AO method, where each subproblem admits either a closed-form or convex solution. Specifically, the receiver-side subproblem is convex for the fixed transmit beamforming set and yields the MMSE receivers, whereas the transmitter-side subproblem becomes a convex quadratic programming for the fixed receive beam set. To deal with this, this subproblem is equivalently transformed into a weighted MSE minimization problem by exploiting the SINR-MMSE relationship from \textbf{Lemma} \textbf{\ref{MMSE_receiver_proof}}. Then, an auxiliary weight variable is introduced to linearize the logarithmic function, yielding a block-wise convex formulation. Theoretically, the optimization variables, objective structures, and constraint forms remain identical across the SS, SA, and SM protocols. Therefore, the AO-WMMSE-based iterative solution process is commonly applicable to all three protocols.

\subsection{Solution of WSR Maximization for SS}
Based on \eqref{min_WSR1} and Table \ref{tab:mse_wsr}, with the fixed transmit beamforming set $\{w_k, v_k\}$ the receiver-side WSR subproblem can be formulated as
\begin{subequations}
    \label{P_1.0a}
    \begin{align}
        \mathbf{P}_{1.0.\text{a}}: \quad & \max_{z, u_k} \sum_{k=1}^{K}\theta_k \log_{2}\left(1+ \text{SINR}^{\text{SS}}_k \right)
        \label{min_WSR_ss0} \\
        \mathrm{s.t.} \quad & \eqref{MSEconstraint1}, \eqref{powerconstraint0},
       \end{align}
\end{subequations}
then, the MMSE receivers can be derived as
\begin{equation}
    \label{z_wsr_ss}
    z^{\text{W-SS}} = \frac{\sum_{k} H_k w_k}{|\sum_{k}H_k w_k|^2 + \sum_{k}|H_k v_k|^2 + \sigma^2},
\end{equation} 
and
\begin{equation}
    \label{u_wsr_SS}
    u_k^{\text{W-SS}} =\frac{H_k v_k}{|\sum_{i}^{K}H_i w_i|^2 + \sum_{i\neq k}|H_i v_i|^2 + \sigma^2}.
\end{equation}

The other transmitter-side WSR subproblem is expressed as
\begin{subequations}
    \label{P_1.0b}
    \begin{align}
        \mathbf{P}_{1.0.\text{b}}: \quad & \max_{w_k, v_k} \sum_{k=1}^{K}\theta_k \log_{2}\left(1+ \text{SINR}^{\text{SS}}_k \right)
        \label{min_WSR_ss1} \\
        \mathrm{s.t.} \quad & \eqref{MSEconstraint1}, \eqref{PAconstraint0}-\eqref{powerconstraint0}.
       \end{align}
\end{subequations}

Based on \textbf{Lemma \ref{MMSE_receiver_proof}}, we can obtain the SINR-MMSE relationship $a_k = \frac{1}{1+\text{SINR}^{\text{SS}}_k}$ from \eqref{a_k}. The \eqref{min_WSR_ss1} can also be written as $\min_{w_k, v_k} \sum_{k=1}^{K}\theta_k \log_{2}\left(a^{\text{SS}_k} \right)$. However, this problem is still nonconvex due to the sum logarithmic structure. We introduce the auxiliary variable $\beta_k$ to represent weight for receive sensing beam and then employ the WMMSE method to transform \eqref{P_1.0b} into
\begin{subequations}
    \label{P_1.0c}
    \begin{align}
        \mathbf{P}_{1.0.\text{c}}: \quad & \min_{w_k, v_k} \sum_{k=1}^{K}\theta_k \left(\beta_k \text{MSE}^{\text{SS}} -\log_2(\beta_k) \right)
        \label{min_WSR_ss2} \\
        \mathrm{s.t.} \quad & \eqref{MSEconstraint1}, \eqref{PAconstraint0}-\eqref{powerconstraint0},
       \end{align}
\end{subequations}
where $\beta_k = 1/\text{MSE}^{\text{SS}}$, and this problem is convex and can be solved by the Karush-Kuhn-Tucker (KKT). The KKT conditions yield the following closed-form solution,
\begin{equation}
    \label{w_wsr_ss}
    w^{\text{W-SS}}_k=\frac{\theta_k\beta_k u_k z H_k}{|z|^2\sum_{i=1}^{K}\theta_i\beta_i|H_i|^2|u_i|}.
\end{equation}

Moreover, the closed-form solution of $v_k$ is given by
\begin{equation}
    \label{v_wsr_ss}
    v_k^{\text{W-SS}} = \sqrt{P^{\max}_k}e^{j \arg(H_k u^{\text{W-SS}}_k)}.
\end{equation}

\subsection{Solution of WSR Maximization for SA}
Under the SA protocol, the WSR maximization problem is also nonconvex, which is similar with that of SS protocol. Therefore, we utilize the same method and algorithm to solve this problem for SA protocol. The receiver-side subproblem is written by
\begin{subequations}
    \label{P_1.1a}
    \begin{align}
        \mathbf{P}_{1.1.\text{a}}: \quad & \max_{z, u_k} \sum_{k=1}^{K}\theta_k \log_{2}\left(1+ \text{SINR}^{\text{SA}}_k \right)
        \label{min_WSR_sa0} \\
        \mathrm{s.t.} \quad & \eqref{MSEconstraint1}, \eqref{PAconstraint0}-\eqref{powerconstraint0},
       \end{align}
\end{subequations}
the corresponding MMSE receivers with receive computation beam is given by
\begin{equation}
    \label{z_wsr_sa}
    z^{\text{W-SA}} = \frac{\sum^{K}_{k} \sum_{m}^{M} H_k w_k}{|\sum^{K}_{k} \sum_{m}^{M} H_k w_k|^2 + \sum^{K}_{k}|\sum^{M}_{m} H_k v_k|^2 + M\sigma^2},
\end{equation} 
the receive sensing beam is expressed as
\begin{equation}
    \label{u_WSR_SA}
    u_k^{\text{W-SA}} =\frac{\sum_{m}^{M} H_k v_k}{|\sum_{i}^{K} \sum_{m}^{M} H_i w_i|^2 + \sum_{i\neq k}^{K}|\sum_{m}^{M} H_i v_i|^2 + M\sigma^2}.
\end{equation}

Further, the transmitter-side WSR subproblem is formulated as
\begin{subequations}
    \label{P_1.1b}
    \begin{align}
        \mathbf{P}_{1.1.\text{b}}: \quad & \max_{w_k, v_k} \sum_{k=1}^{K}\theta_k \log_{2}\left(1+ \text{SINR}^{\text{SA}}_k \right)
        \label{min_WSR_sa1} \\
        \mathrm{s.t.} \quad & \eqref{MSEconstraint1}, \eqref{PAconstraint0}-\eqref{powerconstraint0}.
       \end{align}
\end{subequations}

Likewise, the solution is same as that of SA. We also introduce the auxiliary variable $\tilde{\beta}_k = 1/\text{MSE}^{\text{SA}}$ to represent weight for receive sensing beam and then employ the WMMSE method to transform \eqref{P_1.1b} into
\begin{subequations}
    \label{P_1.1c}
    \begin{align}
        \mathbf{P}_{1.1.\text{c}}: \quad & \min_{w_k, v_k} \sum_{k=1}^{K}\theta_k \left(\tilde{\beta}_k \text{MSE}^{\text{SA}} -\log_2(\tilde{\beta}_k) \right)
        \label{min_WSR_sa2} \\
        \mathrm{s.t.} \quad & \eqref{MSEconstraint1}, \eqref{PAconstraint0}-\eqref{powerconstraint0},
       \end{align}
\end{subequations}
which is convex and can be solved via the KKT conditions. The closed-form solution of transmit beamforming can be derived by,
\begin{equation}
    \label{w_wsr_sa}
    w^{\text{W-SA}}_k=\frac{\theta_k \tilde{\beta}_k u_k z \sum_{m}^{M}H_k}{|z|^2\sum_{i=1}^{K}\theta_i \tilde{\beta}_i|\sum_{m}^{M}H_i|^2|u_i|}.
\end{equation}

Moreover, the closed-form solution of $v_k$ is given by
\begin{equation}
    \label{v_wsr_sa}
    v_k^{\text{W-SA}} = \sqrt{P^{\max}_k}e^{j \arg[(\sum_{m}^{M}H_k)u^{\text{W-SA}}_k]}.
\end{equation}

\subsection{Solution of WSR Maximization for SM}
For SM protocol, the WSR maximization problem is nonconvex and NP-hard. We utilize the AO method to decompose the WSR maximization problem into two subproblems. Different from the SS and SA protocols, the receive beams $\mathbf{z} \in \mathbb{C}^{M \times 1}, \mathbf{u}_k \in \mathbb{C}^{1\times M}$ are vectors. From Table \ref{tab:mse_wsr}, with the fixed transmit beamforming set $w_k, v_k \in \mathbb{C}$ the receiver-side WSR subproblem can be formulated as
\begin{subequations}
    \label{P_1.2a}
    \begin{align}
        \mathbf{P}_{1.2.\text{a}}: \quad & \max_{\mathbf{z}, \mathbf{u}_k}\sum_{k=1}^{K}\theta_k \log_{2}\left(1+ \text{SINR}^{\text{SM}}_k \right)
        \label{min_wsr_sm0} \\
        \mathrm{s.t.} \quad & \eqref{MSEconstraint1}, \eqref{PAconstraint0}-\eqref{powerconstraint0},
       \end{align}
\end{subequations}
the standard MMSE receiver has receive computation beam, which can be derived as
\begin{equation}
    \label{z_wsr_sm}
    \mathbf{z}^{\text{W-SM}} = \frac{\sum_{k} \mathbf{H}_k w_k}{(\sum_{k}\mathbf{H}_k w_k)(\sum_{j}\mathbf{H}_j w_j)^H + \sum_{k}|v_k|^2\mathbf{H}_k \mathbf{H}^H_k + \sigma^2},
\end{equation} 
the receive sensing beam at the BS can be formulated as
\begin{equation}
    \label{u_WSR_SM}
    \mathbf{u}_k^{\text{M-SM}} =\frac{\mathbf{H}_k v_k}{\sum_{i}^{K}|w_i|^2\mathbf{H}_i\mathbf{H}^H_i + \sum_{i\neq k}|v_i|^2\mathbf{H}_i\mathbf{H}^H_i + \sigma^2}.
\end{equation}

The transmitter-side WSR subproblem is formulated as
\begin{subequations}
    \label{P_1.2b}
    \begin{align}
        \mathbf{P}_{1.2.\text{b}}: \quad & \max_{w_k, v_k} \sum_{k=1}^{K}\theta_k \log_{2}\left(1+ \text{SINR}^{\text{SM}}_k \right)
        \label{min_WSR_sm1} \\
        \mathrm{s.t.} \quad & \eqref{MSEconstraint1}, \eqref{PAconstraint0}-\eqref{powerconstraint0}.
       \end{align}
\end{subequations}

Let the auxiliary variable $\hat{\beta}_k = 1/\text{MSE}^{\text{SM}}$ as the weight for receive sensing beam. The WMMSE method is adopted to transform \eqref{P_1.2b} into
\begin{subequations}
    \label{P_1.2c}
    \begin{align}
        \mathbf{P}_{1.2.\text{c}}: \quad & \min_{w_k, v_k} \sum_{k=1}^{K}\theta_k \left(\hat{\beta}_k \text{MSE}^{\text{SM}} -\log_2(\hat{\beta}_k) \right)
        \label{min_WSR_sm2} \\
        \mathrm{s.t.} \quad & \eqref{MSEconstraint1}, \eqref{PAconstraint0}-\eqref{powerconstraint0},
       \end{align}
\end{subequations}
which is convex. Under the KKT conditions, the closed-form solution of transmit computation beamforming is derived by
\begin{equation}
    \label{w_wsr_sm}
    w^{\text{W-SM}}_k = \frac{1 + \sqrt{\zeta}}{\frac{\sum_i^K|\mathbf{z}^H\mathbf{H}_i|^2}{\sum_{k}^{K}\theta_k}\hat{\beta}_k|\mathbf{u}_k\mathbf{H}_k|^2}\cdot\frac{\mathbf{z}^H \mathbf{H}_k}{\sum_k^K \theta_k \hat{\beta}_k\left|\mathbf{u}_k \mathbf{H}_k\right|^2}.
\end{equation}

Moreover, the closed-form solution of $v_k$ is given by 
\begin{equation}
    \label{v_wsr_sm}
    v_k^{\text{W-SM}} = \sqrt{P^{\max}_k}e^{j \arg(\mathbf{H}_k \mathbf{u}^{\text{W-SM}}_k)}.
\end{equation}

The AO-WMMSE-based algorithm for SS, SA, and SM is shown as Algorithm \ref{alg:ao_mmse}.
\begin{algorithm}[htbp]
    \footnotesize
    \caption{The AO-WMMSE-Based WSR Maximization Algorithm for SS, SA, and SM Protocols}
    \label{alg:ao_wmmse}
    \begin{algorithmic}[1]
    \REQUIRE Protocol $P\in\{\mathsf{SS},\mathsf{SA},\mathsf{SM}\}$, $\theta_k$, $\sigma^2$, $P_k^{\max}$, $z$, $w_k$, $v_k$, $u_k$, $T$
    \ENSURE $\{w_k,v_k\}$, $z$, $u_k$ (SS/SA), $\mathbf{z}$, $\{\mathbf{u}_k\}$ for SM.
    \STATE Initialize a feasible $z^{(0)},\{w_k^{(0)}\},\{v_k^{(0)}\}$; $t\!\leftarrow\!0$
    \REPEAT
    \STATE \textbf{Receiver-side subproblem:}
    \IF{$\mathsf{P}=\mathsf{SS}$}
        \STATE Update $z^{(t)}$ and $u_k^{(t)}$ from the MMSE formulas from \eqref{z_wsr_ss} and \eqref{u_wsr_SS} with fixed $\{w_k^{(t-1)},v_k^{(t-1)}\}$.
    \ELSIF{$\mathsf{P}=\mathsf{SA}$}
        \STATE Update $z^{(t)}$ and $u_k^{(t)}$ from \eqref{z_wsr_sa} and \eqref{u_WSR_SA} with fixed $\{w_k^{(t-1)},v_k^{(t-1)}\}$.
    \ELSE
        \STATE Update $\mathbf{u}^{(t)}$ and $\mathbf{u}_k^{(t)}$ from \eqref{z_wsr_sm} and \eqref{u_WSR_SM} with fixed $\{w_k^{(t-1)},v_k^{(t-1)}\}$.
    \ENDIF
    \STATE Compute $a_k^{(t)}$ and update weights $\beta_k^{(t)}=(a_k^{(t)})^{-1}$.
  
    \STATE \textbf{Transmitter-side subproblem:}
    \IF{$\mathsf{P}=\mathsf{SS}$}
        \STATE Update $\{w_k^{(t)},v_k^{(t)}\}$ by the closed-form WMMSE solution by solving \eqref{P_1.0c} from \eqref{w_wsr_ss} and \eqref{v_wsr_ss}
        \ELSIF{$\mathsf{P}=\mathsf{SA}$}
            \STATE Update $\{w_k^{(t)},v_k^{(t)}\}$ by solving \eqref{P_1.1c} from \eqref{w_wsr_sa} and \eqref{v_wsr_sa}
        \ELSE
            \STATE Update $\{w_k^{(t)},v_k^{(t)}\}$ by solving \eqref{P_1.2c} from \eqref{w_wsr_sm} and \eqref{v_wsr_sm}
        \ENDIF
      \STATE $t\!\leftarrow\!t+1$
    \UNTIL{relative decrease of WMMSE objective $\le \varepsilon$}
    \end{algorithmic}
    \end{algorithm}
    
\subsection{Computational Complexity Analysis}

For the SS protocol, all optimization variables are scalars, and the update operations are decoupled among users. During each iteration, computing the MMSE receiver $u_k$ and the weight factor $\beta_k$ involves the evaluation of the common term $\sum_{k}^{K}|w_k|^2$ and several scalar multiplications, which requires a computational cost of $O(K)$. The transmit-side update also requires $O(K)$. The update of the scalar variable $z$ only involves simple projection operations with a constant cost. Therefore, the computational complexity of each iteration is $O(K)$, and the overall computational complexity under the SS protocol is $O(TK)$, where $T$ is the total number of iterations.

For the SA protocol, the aggregated equivalent channel is computed once before the iterative procedure, with an initial cost of $O(KM)$. After channel aggregation, the iterative computational procedure is identical to that of the SS protocol. Specifically, the receiver-side and weight updates cost $O(K)$, and the transmitter-side updates require $O(K)$. The scalar update of $z$ has negligible complexity. Hence, the computational complexity of each iteration under the SA protocol remains $O(K)$, and the overall computational complexity is $O(KM + TK)$, where $O(KM)$ represents the one-time equivalent channel aggregation that does not scale with the number of iterations. 

For the SM protocol, the receiver-side updates involve vector operations, which significantly increase the computational load. Specifically, updating the MMSE receivers $\mathbf{z}\in\mathbb{C}^{M\times1}$ and $\mathbf{u}_k$ requires the construction and inversion of an $M\times M$ covariance matrix, leading to a computational cost of $O(M^3)$. Additionally, each user's receiver update involves matrix-vector products with complexity $O(KM^2)$. The transmitter-side updates require computing costs $O(KM)$, followed by the closed-form update with $O(K)$ operations. Consequently, the computational complexity of each iteration under the SM protocol is dominated by the receiver-side operations, resulting in $O(M^3 + KM^2)$, and the overall computational complexity is $O(T(M^3 + KM^2))$.

\section{Simulation Analysis}
In this simulation, we conduct numerical simulations to evaluate the effectiveness of the proposed JCC-PASS framework. Unless otherwise specified, the parameters are set as follows. We assume that UEs are randomly distributed within the rectangular area $20 m \times 20 m$ along $x$-axis and $y$-axis, where $D^{\text{y}}_0 = 10 $m, $D^{\text{y}}_{\max}=20 $m and $D^{\text{x}}_0 = 20 $m. The waveguide array is deployed at a fixed height $D^{\text{y}}_0 = 3$m. The first waveguide feed point is located at $[0, 10, 3]$m. The carrier frequency is $f_c = 28 $GHz, and the effective refractive index of the dielectric waveguide is $n_{\text{eff}} = 1.4$. The average in-waveguide attenuation factor is set as $\kappa = 0.08 $dB/m and $\alpha = 0.0092 m^{-1}$. The inter-antenna spacing is configured as $\Delta = \lambda/2$, and the maximum transmit power of each UE is denoted by $P^{\max}_k = 10 $dBm, $\forall k \in [1,32]$, while the noise power is $\sigma^2 = -90 $dBm. For computational performance evaluation, we adopt the normalized MSE per user, i.e., $\text{MSE}/K$, as the performance metric, and all results are averaged over 1000 independent channel realizations. The benchmark schemes include the conventional MIMO, conventional PASS employing a single continuous waveguide and the proposed JCC-PASS framework for Case I without in-waveguide loss and Case II with in-waveguide loss.

\begin{figure}[htbp]
    \centering
    \begin{subfloat}[]{\includegraphics[width=3.4in]{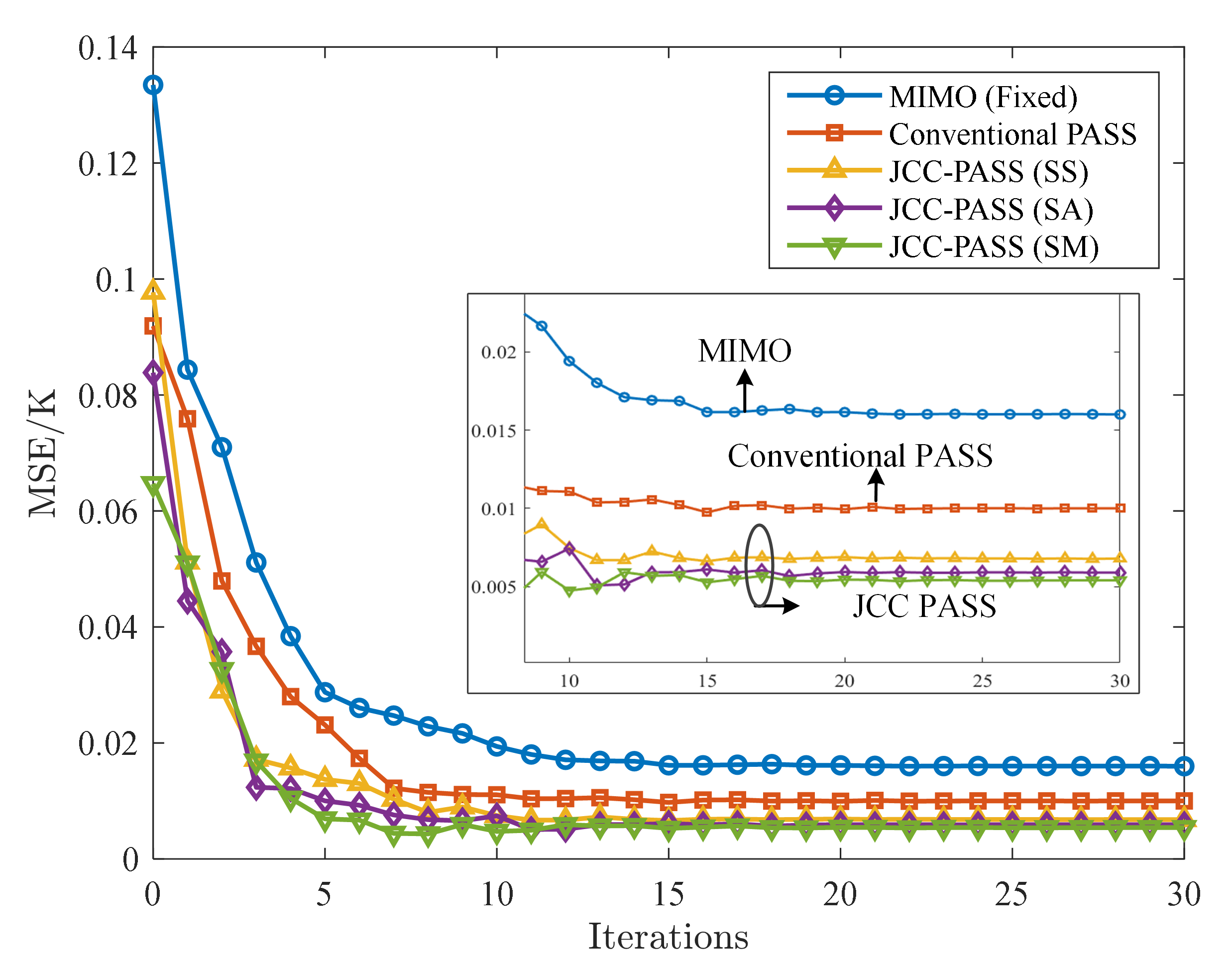}
    \label{fig_2a}}
    \end{subfloat}
    \centering
    \begin{subfloat}[]{\includegraphics[width=3.4in]{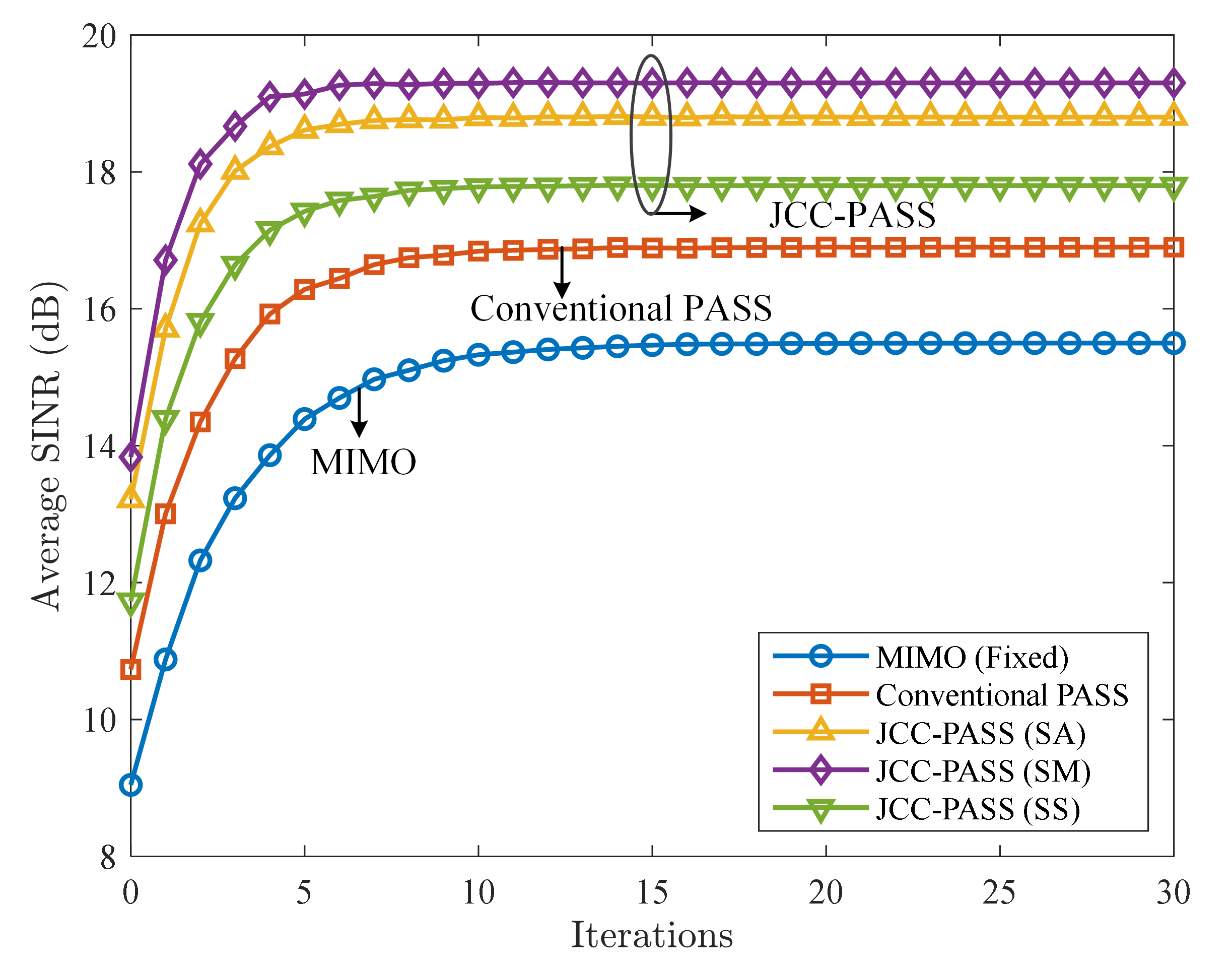}
    \label{fig_2b}}
    \end{subfloat}
    \caption{Comparisons of convergence behaviors of the proposed algorithms under SS, SA, and SM protocols. (a) convergence versus iterations of AO-MMSE. (b) convergence versus iterations of AO-WMMSE.}
    \label{fig_2}
  \end{figure}

  Fig.~\ref{fig_2} illustrates the convergence behaviors of the proposed JCC-PASS framework in terms of the MSE/K and the average SINR versus the number of iterations. It can be observed that all curves in Fig.~\ref{fig_2a} exhibit a rapid decrease in MSE during the first several iterations and gradually converge to stable values after approximately 10-15 iterations, confirming the fast convergence and robustness of the proposed AO-MMSE-based algorithm. The conventional PASS reduces the estimation error by nearly 40$\%$ compared with MIMO, demonstrating the benefit of reconfigurable PA structures. The JCC-PASS framework under the SM protocol achieves the lowest final MSE, outperforming the conventional PASS by over 70$\%$. As shown in the Fig.~\ref{fig_2b}, all frameworks deployed the AO-WMMSE-based algorithm exhibit stable convergence within approximately 15-20 iterations, demonstrating the efficiency of the proposed AO-WMMSE algorithm. The conventional PASS achieves about a 2 dB improvement over MIMO, benefiting from the flexible PA configuration. The proposed JCC-PASS framework substantially enhances the SINR performance by jointly exploiting segment-level coordination. Specifically, the SS protocol adaptively activates the most efficient waveguide segments, achieving roughly 1.5 dB gain over the conventional PASS. The SA protocol aggregates multiple adjacent segments to improve beam coherence and energy concentration, leading to a further 2.5 dB gain. The SM protocol, which allows simultaneous multiplexing across multiple active segments, attains the highest steady-state SINR of approximately 19.8 dB, offering a 3.5 dB advantage over the conventional PASS and exhibiting the fastest convergence speed.

  \begin{figure}[htbp]
    \centering
    \includegraphics[width=3.4in]{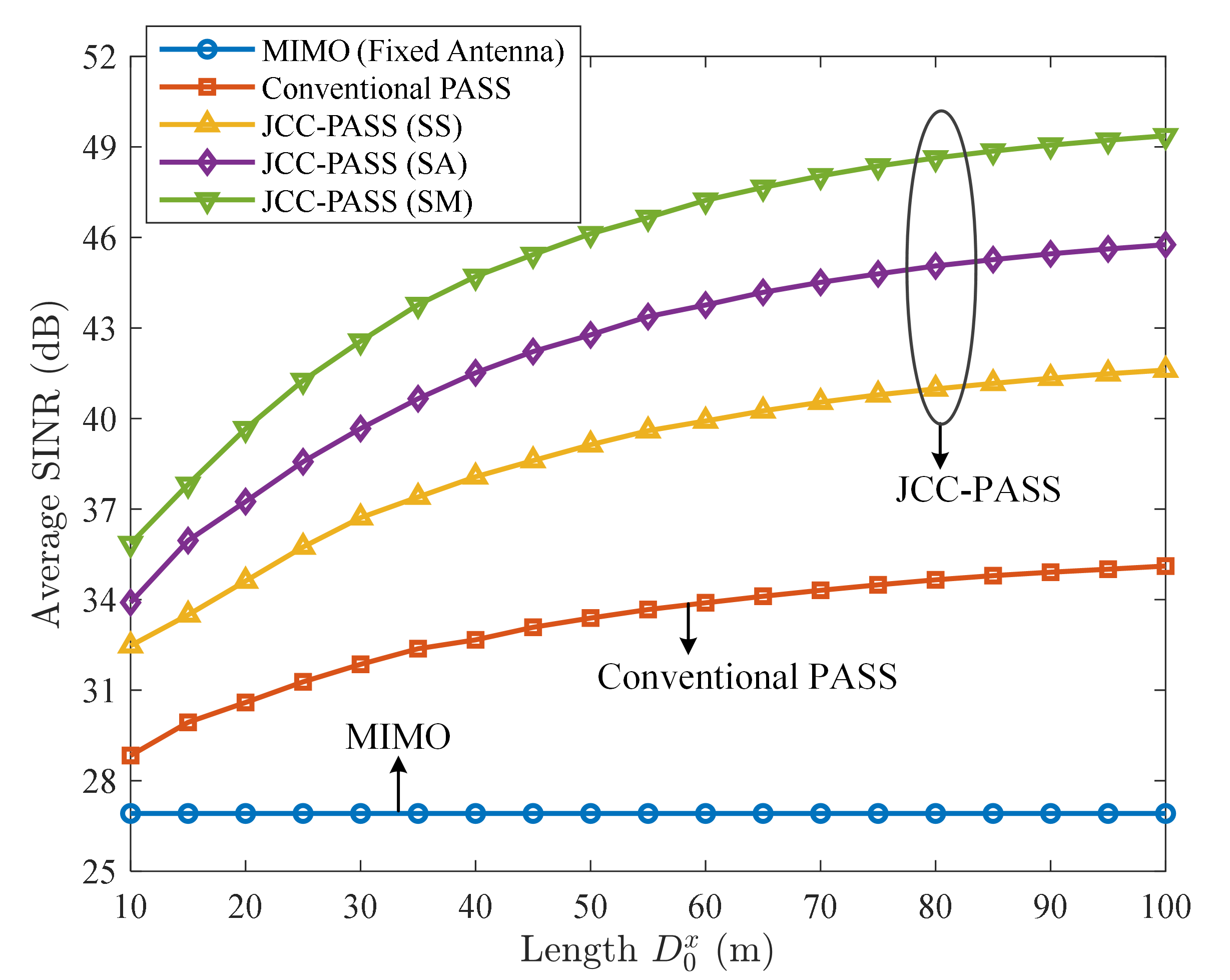}
    \caption{Comparisons of the average SINR versus different lengths of the waveguide.}
    \label{fig_3}
  \end{figure}

  Fig.~\ref{fig_3} illustrates the variation of the average SINR with respect to the waveguide length $D^\text{x}_0$ for different frameworks. Quantitatively, when the waveguide length reaches 100 m, the proposed JCC-PASS architecture yields remarkable SINR improvements compared with the baselines.
  Specifically, the JCC-PASS under SM protocol achieves approximately 86.14$\%$ and 72.21$\%$ SINR gains over the conventional MIMO and PASS systems, respectively. These substantial gains verify that segment-level reconfigurability enables JCC-PASS to efficiently utilize extended waveguide apertures for interference suppression and beamforming enhancement, demonstrating a clear performance scalability advantage over both conventional antenna and single-waveguide configurations.

  \begin{figure}[htbp]
    \centering
    \includegraphics[width=3.4in]{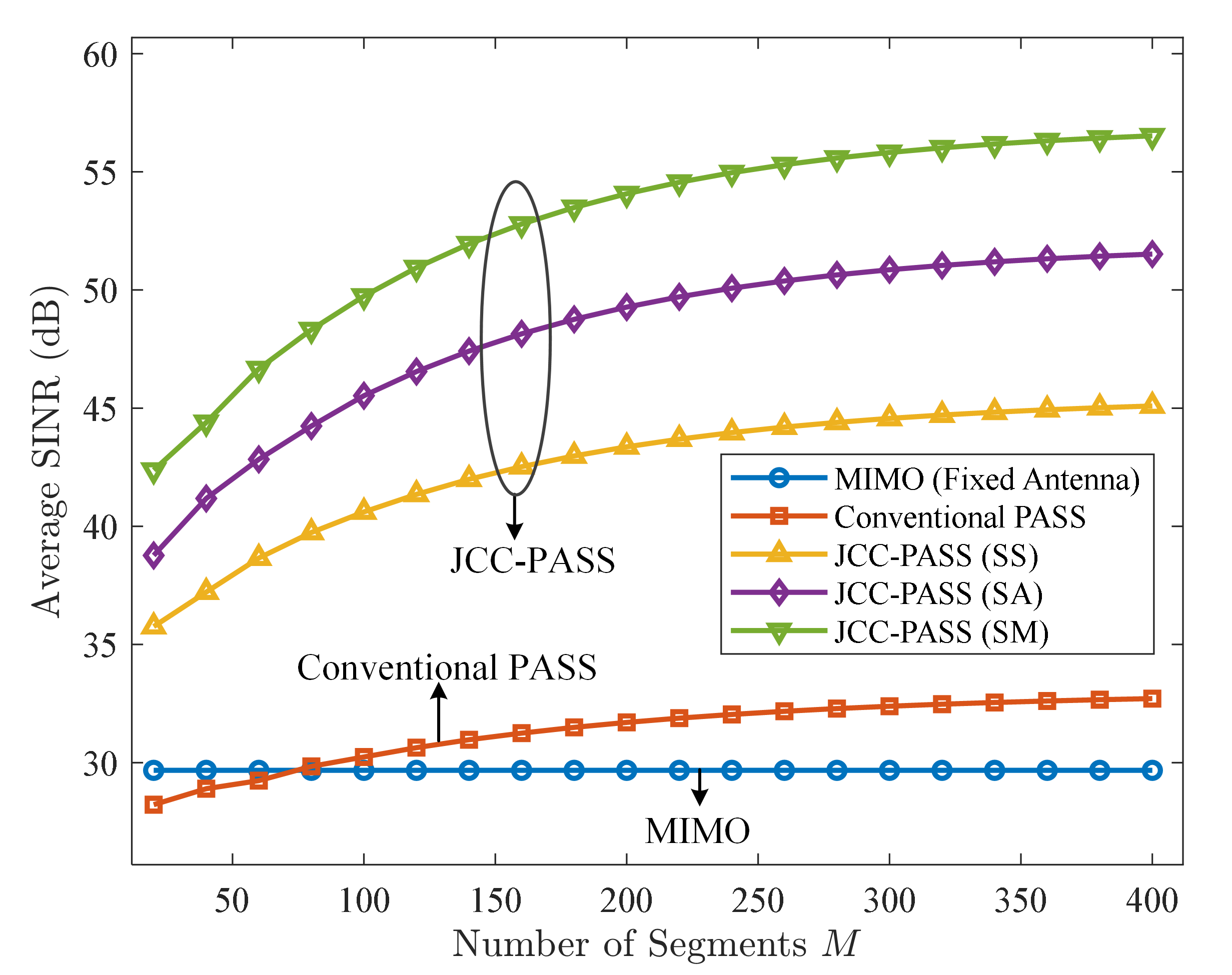}
    \caption{Comparisons of the average SINR versus different number of segments.}
    \label{fig_4}
  \end{figure}

  Fig.~\ref{fig_4} shows the effect of the number of waveguide segments $M$ on the average SINR performance. As the number of segments increases, all PASS-based systems experience noticeable SINR improvements, since finer segmentation provides higher degrees of freedom for pinching control and more accurate beam steering. In contrast, the conventional MIMO system remains nearly constant at around 30 dB, as its antenna structure is fixed and cannot exploit segment-level flexibility. At $M=400$, the proposed JCC-PASS framework under SS, SA, and SM protocols achieve remarkable SINR gains compared with benchmark systems. Specifically, the JCC-PASS under SM provides approximately 96$\%$, 79.01$\%$ improvements over the conventional MIMO system and the conventional PASS architecture.

  \begin{figure}[htbp]
    \centering
    \includegraphics[width=3.4in]{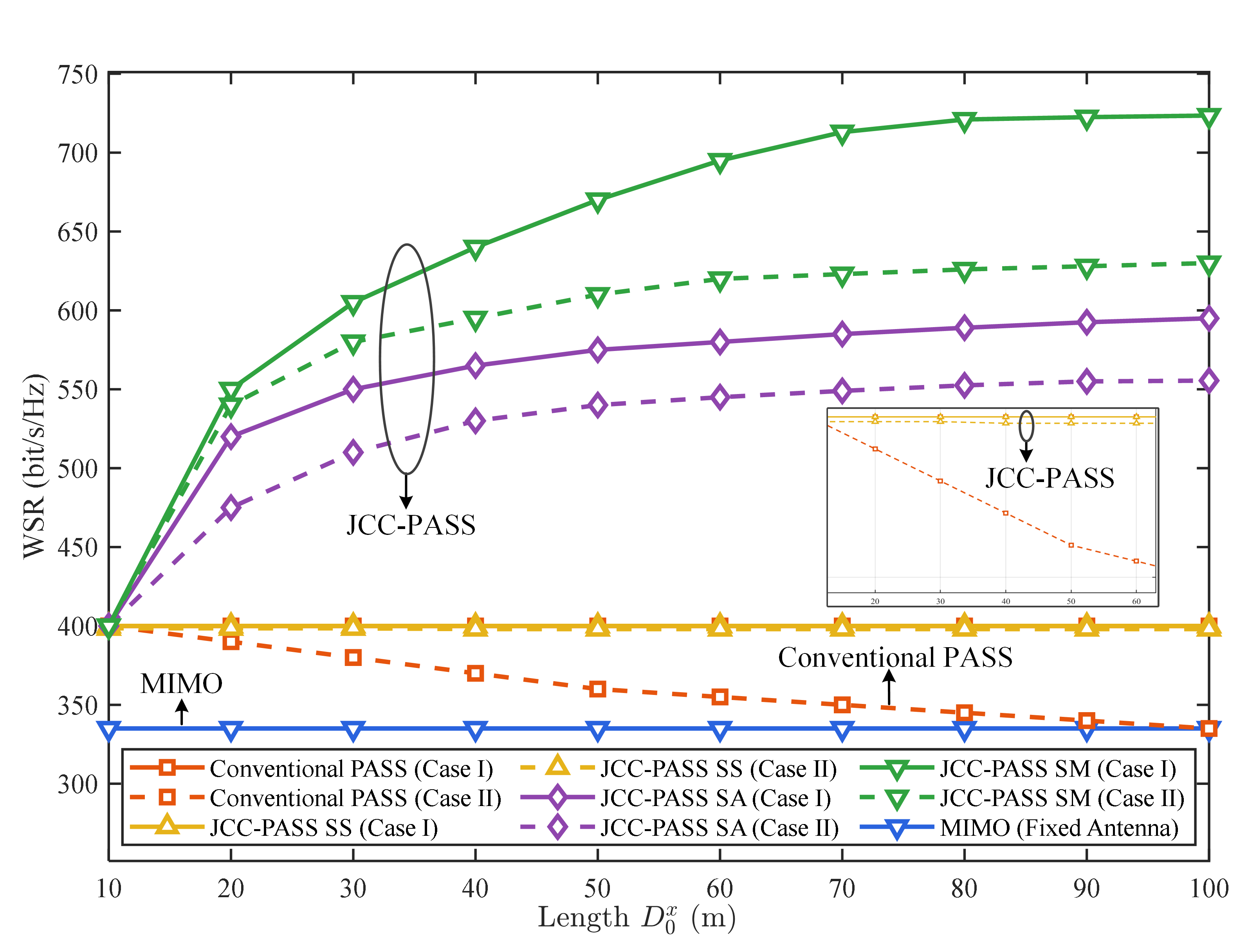}
    \caption{Comparisons of the WSR versus different lengths of the waveguide for Case I and Case II.}
    \label{fig_5}
  \end{figure}

  Fig.~\ref{fig_5} illustrates the variation of the WSR versus the waveguide length $D^x_0$ for different architectures for case I and case II. As observed, across all three operation protocols, the proposed JCC-PASS consistently outperforms both the conventional PASS and the fixed-antenna MIMO architectures, regardless of the waveguide length. At $D^x_0$=100 m, the JCC-PASS frameworks under the SS, SA, and SM protocols achieve approximately 200 bit/s/Hz, 250 bit/s/Hz, and 290 bit/s/Hz, these correspond to 30.02$\%$, 39.50$\%$, and 51.35$\%$ higher WSRs than the conventional PASS, 82.12$\%$, 84.43$\%$, and 87.70$\%$ gains compared with the conventional MIMO.
  The performance advantage becomes more pronounced under Case II, where the effect of dielectric waveguide attenuation is included. Even with in-waveguide propagation loss, the proposed JCC-PASS maintains a clear WSR superiority, indicating its robustness to physical channel degradation. In contrast, the conventional PASS suffers an evident rate drop as the waveguide length increases due to accumulated propagation loss and limited reconfigurability.
  Among the proposed variants, JCC-PASS (SM) consistently achieves the highest WSR owing to its ability to multiplex multiple active segments simultaneously, which enhances spatial parallelism and reduces interference. The SA protocol also provides significant gain by coherently combining adjacent segment signals, whereas the SS protocol offers a lightweight design with lower computational overhead but relatively smaller gain.

  \begin{figure}[htbp]
    \centering
    \includegraphics[width=3.4in]{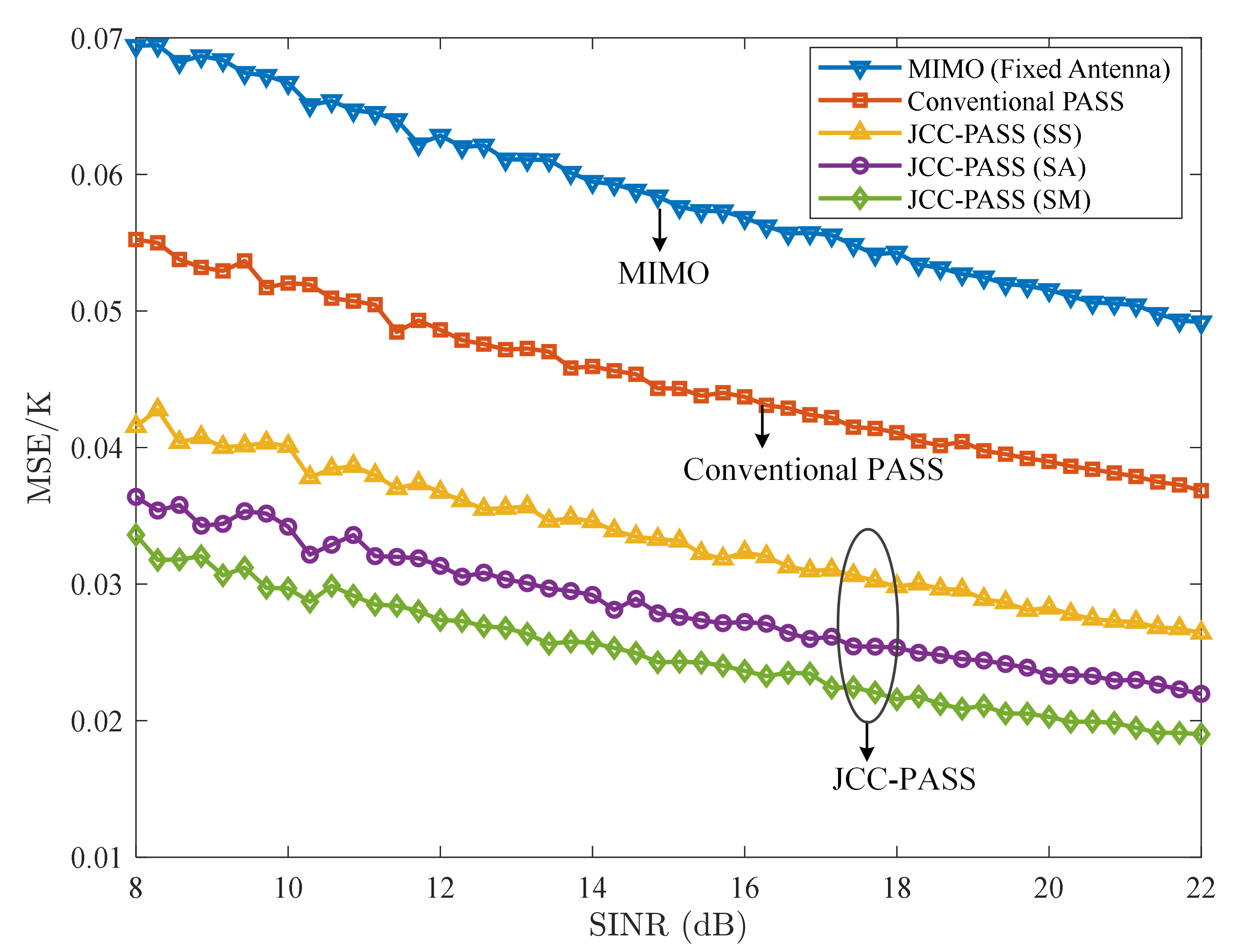}
    \caption{Comparisons of the MSE/K versus the receive SINR.}
    \label{fig_6}
  \end{figure}

  Fig.~\ref{fig_6} presents the relationship between the MSE/K and the receive SINR for different frameworks. The MSE of all schemes decreases monotonically with increasing SINR, verifying that higher signal quality leads to more accurate channel estimation and data reconstruction. The conventional PASS moderately improves the estimation accuracy, reducing the steady-state MSE by about 28$\%$ compared with the conventional MIMO. The SM protocol provides the best performance, achieving 45.32$\%$ lower MSE compared with the conventional PASS and 70.65$\%$ reduction compared with MIMO, demonstrating the effectiveness of multi-segment multiplexing and joint communication-computation optimization.
  Moreover, the slope of the MSE curve under the JCC-PASS (SM) and JCC-PASS (SA) schemes is steeper, indicating that these protocols benefit more from SINR improvement and exhibit stronger robustness under high-SNR regimes. This demonstrates that multi-segment multiplexing and cooperative aggregation can effectively mitigate residual estimation error and enhance convergence stability. Overall, the results validate the superiority of the proposed JCC-PASS framework in achieving reliable signal recovery and efficient communication-computation coupling across a wide SINR range.

\section{Conclusion}
In this paper, an uplink JCC-PASS framework enabled by segment-waveguide method has been proposed to achieve the computation-communication trade-off by simultaneously transmitting computation data and communication bit streams. The computation-oriented MSE minimization and communication-oriented WSR maximization problems have been formulated to jointly optimize the transmit beamforming at UEs and the receive beams at the BS. To tackle the strong coupling and nonconvexity between variables, two low-complexity iterative algorithms, namely the AO-MMSE-based and AO-WMMSE-based algorithms, have been developed to obtain closed-form solutions. In particular, the AO-MMSE-based algorithm has jointly optimized the receiver and transmitter through alternating updates, where the receiver-side MSE subproblem has been minimized via MMSE filtering and then the transmitter-side subproblem has been reformulated into a convex SDR form for efficient optimization. It has been mathematically proved that the proposed JCC-PASS framework outperforms conventional PASS in the in-waveguide gain, which is more pronounced as segment or waveguide length become greater. For the communication-oriented WSR maximization problem, the AO-WMMSE-based algorithm has introduced an auxiliary weighting variable to linearize the logarithmic function and iteratively updated both the transmit beamforming and receive beams. Extensive simulations have demonstrated that the proposed JCC-PASS framework can achieve up to 70.65$\%$ and 45.32$\%$ reduction in MSE compared with conventional MIMO and conventional PASS, it also can obtain 87.70$\%$ and 51.35$\%$ improvements in WSR compared with these two systems. Future work will focus on extending the proposed framework toward learning-assisted control and distributed optimization to support real-time adaptation under dynamic channel conditions.



\bibliographystyle{IEEEtran}
\def\baselinestretch{1}
\bibliography{JCC_PASS}

\end{document}